\documentclass[ALICE,manyauthors]{cernphprep}
\usepackage[english]{babel}
\usepackage{graphicx}
\usepackage{verbatim}
\usepackage{amsfonts}
\usepackage{makeidx}
\usepackage{color}
\usepackage{amsmath}
\usepackage{lineno}
\usepackage{epstopdf}
\usepackage{hyperref}
\usepackage{cite}
\newcommand{\pp}{{\rm pp}}
\newcommand{\sqrts}{\sqrt{s}}

\newcommand{\gsim}{\,{\buildrel > \over {_\sim}}\,}
\newcommand{\av}[1]{\left\langle #1 \right\rangle}

\newcommand{\MeV}{\mathrm{MeV}}
\newcommand{\GeV}{\mathrm{GeV}}
\newcommand{\TeV}{\mathrm{TeV}}

\newcommand{\gev}{\mathrm{GeV}}
\newcommand{\tev}{\mathrm{TeV}}

\newcommand{\mum}{\mathrm{\mu m}}

\newcommand{\mb}{\mathrm{mb}}
\newcommand{\mub}{\mathrm{\mu b}}

\newcommand{\pt}{\ensuremath{p_{\rm T}}}
\newcommand{\kt}{\ensuremath{k_{\rm T}}}
\newcommand{\mt}{\ensuremath{m_{\rm T}}}

\newcommand{\Dstophipi}{{\rm D_s^{+}\to \phi\pi^+}}
\newcommand{\phitoKK}{{\rm \phi\to  K^-K^+}}
\newcommand{\DstoKzerostarK}{{\rm D_s^{+}\to \overline{K}^{*0} K^+}}
\newcommand{\Dstofzeropi}{{\rm D_s^{+}\to f_{0}(980) \pi^+}}
\newcommand{\fzero}{{\rm f_{0}(980)}}
\newcommand{\Kzerostar}{{\rm \overline{K}^{*0}}}
\newcommand{\phitoKKpi}{{\rm \phi\to K^-K^+}}
\newcommand{\DstophipitoKKpi}{{\rm D_s^{+}\to \phi\pi^+\to K^-K^+\pi^+}}
\newcommand{\DstoKzerostarKtoKKpi}{{\rm D_s^{+}\to \overline{K}^{*0} K^+ \to K^-K^+\pi^+ }}
\newcommand{\KKpi}{{\rm K^-K^+\pi^+}}
\newcommand{\Dzero}{{\rm D^0}}
\newcommand{\Dstar}{{\rm D^{*+}}}
\newcommand{\Dplus}{{\rm D^+}}
\newcommand{\Ds}{{\rm D_s^+}}
\newcommand{\Dsplusminus}{{\rm D_s^{\pm}}}
\newcommand{\nbinv}{{\rm nb^{-1}}}

\newcommand{\dEdx}{{\rm d}E/{\rm d}x}
\newcommand{\gammas}{\gamma_{\rm s}}
\newcommand{\epluseminus}{{\rm e^+e^-}}
\begin{document}

\begin{titlepage}

\PHnumber{2012-227}                 % required, obtained from PH
\PHdate{09 Aug 2012}              % required

\title{$\Ds$ meson production at central rapidity \\in proton--proton collisions at $\sqrts = 7~\TeV$}

\Collaboration{ALICE Collaboration%
         \thanks{See Appendix~\ref{app:collab} for the list of collaboration 
                      members}}
\ShortAuthor{ALICE Collaboration}
\ShortTitle{$\Ds$ meson production at central rapidity 
in pp collisions at $\sqrts = 7~\TeV$}

\begin{abstract}
The $\pt$-differential inclusive production cross section of the prompt 
charm-strange meson $\Ds$ 
in the rapidity range $|y|<0.5$ was measured in proton--proton collisions at 
$\sqrt{s}=7~\tev$ at the LHC using the \mbox{ALICE} detector. 
The analysis was performed on a data sample of 
$2.98 \times 10^8$ events collected with a minimum-bias trigger.
The corresponding integrated luminosity is $L_{\rm int}=4.8~\nbinv$. 
Reconstructing the decay $\Dstophipi$, with $\phitoKK$, and its charge 
conjugate, about 480 $\Dsplusminus$ mesons were counted, after selection cuts, 
in the transverse momentum range $2<\pt<12~\gev/c$.
The results are compared with predictions from models based on perturbative 
QCD. 
The ratios of the cross sections of four D meson species (namely 
$\Dzero$, $\Dplus$, $\Dstar$ and $\Ds$) were determined both
as a function of $\pt$ and integrated over $\pt$ 
after extrapolating to full $\pt$ range,
together with the strangeness suppression factor in charm 
fragmentation.
The obtained values are found to be compatible within uncertainties 
with those measured by other experiments
in $\epluseminus$, ep and pp interactions at various centre-of-mass energies.
\end{abstract}
\end{titlepage}
\setcounter{page}{2}
%\linenumbers

%%
\section{Introduction}
\label{sec:intro}
The measurement of open charm production in proton--proton ($\pp$) collisions 
at the Large Hadron Collider (LHC) provides a way to test predictions
of quantum chromodynamics (QCD) at the highest available
collision energies.
Charm and beauty production cross sections can be computed 
in perturbative QCD (pQCD) using the 
factorization approach~\cite{FONLL,GMVFNS}.
In this scheme, cross sections are computed as a convolution of three terms:
the parton distribution functions of the incoming protons, the partonic hard 
scattering cross section, and the fragmentation process.
The partonic hard scattering cross section is computed through a 
perturbative calculation~\cite{FONLL,GMVFNS}, while the
parton distribution functions and the fragmentation process are parametrized on 
experimental data.
In particular, the fragmentation describes the non-perturbative 
transition of a charm quark to a hadron.
It is modeled by a fragmentation function, which parametrizes the 
fraction of quark energy transferred to the produced hadron, and by the 
fragmentation fractions, $f({\rm c\rightarrow D})$, which describe the 
probability of a 
charm quark to hadronize into a particular hadron species. 

The production of prompt $\Dzero$, $\Dplus$ and $\Dstar$ mesons in $\pp$ 
collisions at $\sqrts=7~\tev$ was measured with the ALICE 
detector at two centre-of-mass energies, namely 7 and 
2.76~$\TeV$~\cite{alicecharm,alicecharmlowen}.
Here, `prompt' indicates D mesons produced at the $\pp$ interaction point, 
either directly in the hadronization of the charm quark or in
strong decays of excited charm resonances.
The contribution from weak decays of beauty mesons, which
give rise to feed-down D mesons displaced from the interaction vertex,
was subtracted.
The measured $\pt$-differential cross sections for prompt $\Dzero$, $\Dplus$ 
and $\Dstar$ are described within uncertainties by theoretical predictions 
based on pQCD at next-to-leading order (e.g. in the general-mass 
variable-flavour-number scheme, GM-VFNS~\cite{VFNSalice}) or 
at fixed order with next-to-leading-log resummation (FONLL~\cite{FONLLalice}).
The central value of the GM-VFNS predictions for these three mesons
lies systematically above the data.
On the other hand, the data tend to be higher than the central value of the 
FONLL 
predictions, as it was observed at lower collision energies, namely at the
Tevatron~\cite{cdf,fonllDcdf}, where hadronic decays of D mesons were
reconstructed, and at RHIC, where measurements of electrons from semileptonic 
D and B decays were performed~\cite{phenix,star}.

Heavy flavour production in hadronic collisions can be calculated also
in the framework of $\kt$-fac\-toriza\-tion with
unintegrated gluon distributions (UGDFs) to account for 
the transverse momenta of the initial 
partons~\cite{Catani,Collins,SmallX,MSRHIC}.
Calculations of inclusive production cross section of D mesons
based on this approach in the leading order (LO) 
approximation were recently published for LHC energy
and compared to experimental results~\cite{MS1,MS2}. 

The measurement of the $\pt$-differential prompt $\Ds$ meson production
is of particular interest due to its strange valence quark content.
The $\Ds$ production cross section in hadronic collisions
was measured at lower energies at the Tevatron collider
in the transverse momentum ($\pt$) range $8 < \pt < 12~\gev/c$~\cite{cdf}.
Preliminary results for $\Ds$ production at the LHC were reported by the
LHCb Collaboration for prompt mesons at forward rapidity~\cite{LHCbDnote} and 
by the ATLAS Collaboration at central rapidity~\cite{ATLASDnote}.
The LHCb Collaboration also measured the asymmetry between
prompt $\rm D_s^+$ and $\rm D_s^-$ production in the rapidity region 
$2<y<4.5$ and for transverse momenta $\pt>2~\GeV/c$, observing a small excess
of $\rm D_s^-$ mesons:
$A_{\rm P}=(\sigma(\rm D_s^+)-\sigma(\rm D_s^-))/(\sigma(\rm D_s^+)+\sigma(\rm D_s^-))=(-0.33\pm0.22\pm0.10)$\%~\cite{LHCbAsymm}.
Such a particle-antiparticle production asymmetry is understood in 
phenomenological models as due to the effect of the beam remnants on the 
heavy-quark hadronization, see e.g.~\cite{Norrbin}.

Charm production has been measured in ep interactions at the HERA 
collider by the ZEUS~\cite{Zeus} and H1~\cite{h1} Collaborations, as well 
as in $\epluseminus$ annihilations, 
at the Z$^{0}$ resonance, by the ALEPH~\cite{ALEPH}, DELPHI~\cite{DELPHI} and 
OPAL~\cite{OPAL} Collaborations, and at centre-of-mass energies of about 10 GeV
by the CLEO~\cite{CLEO} and ARGUS~\cite{ARGUS} Collaborations.

As far as theoretical models are concerned, a calculation of the $\Ds$ 
production cross section within the FONLL framework
is not available, because of the poor knowledge of the parton fragmentation 
function.
The measured data points can be compared with the GM-VFNS prediction 
that uses meson specific fragmentation functions~\cite{BAKFF1}.

From the differential production cross section of prompt $\Dzero$, 
$\Dplus$, $\Dstar$ and $\Ds$ mesons, the relative production yields of the
D meson species can be studied as a function of transverse momentum. 
A $\pt$ dependence is expected for these ratios, due to differences in the 
fragmentation function of the charm quark in the four 
considered meson species, and because of the different 
contributions from decays of higher excited states.
In this sense, the measurement of the ratios between the D meson species
can provide information 
on the fragmentation functions that can be used in the pQCD models based on the 
factorization approach.
%The relative abundances of charm mesons are determined by the charm 
%fragmentation fractions, $f({\rm c\rightarrow D})$.
The suppression of strange meson production in the charm fragmentation is
quantified by the strangeness suppression factor, $\gamma_{\rm s}$, which is
computed from the measured  $\Dzero$, $\Dplus$ and $\Ds$ 
cross sections extrapolated to full $\pt$ range, as defined in 
Section~\ref{sec:results}.
The values measured at the LHC can be compared with those measured for different
energies and different colliding systems~\cite{Zeus2}.

Furthermore, the measurement of $\Ds$ in $\pp$ collisions provides a reference
for the studies of charm production in heavy-ion collisions.
According to QCD calculations on the lattice,
under the conditions of high energy-density and temperature that are
reached in these collisions, the confinement of quarks and gluons into hadrons 
vanishes and a transition to a Quark-Gluon Plasma (QGP) occurs~\cite{QGP}.
Charm hadrons are a powerful tool to study the properties of the QCD medium 
created in these collisions~\cite{dk,DinQGP,aliceDRAA}.
In particular, the $\Ds$ meson is sensitive to strangeness production 
in heavy-ion collisions.
Strange quarks are abundant in the QGP, resulting in an 
enhanced production of strange particles with respect to pp 
collisions~\cite{na57,STARstrange1,STARstrange2,ALICEstrange}.
Hence, at low momentum, the relative yield of $\Ds$ mesons with respect to 
non-strange charm mesons 
(such as $\Dzero$ and $\Dplus$) is predicted to be enhanced in 
nucleus-nucleus collisions~\cite{rafelski,HeFriesRapp,Andronic}, if the 
dominant mechanism for D meson formation at low/intermediate momenta is 
in-medium hadronization of charm quarks via coalescence with strange quarks
~\cite{starexp1,starexp2,phenix2}.

In this paper, we report on the measurement of $\Ds$ production
cross section in $\pp$ collisions at \mbox{$\sqrts=7~\TeV$} with the 
ALICE detector at the LHC. 
$\Ds$ mesons were reconstructed through their hadronic decay channel 
$\Dstophipi$ with a subsequent decay $\phitoKK$.
The $\pt$-differential cross section is measured over a range
of transverse momentum extending from 2~$\gev/c$ up to 12~$\gev/c$ at
central rapidity, $|y|<0.5$.  
In Section~\ref{sec:detector}, the detector layout and the 
data sample are described. 
This is followed, in Section~\ref{sec:sele}, by the description of
the $\Ds$ meson reconstruction strategy, the selection cuts, and the
raw yield extraction from the invariant mass distributions.
The  various corrections applied to obtain the production cross sections 
are illustrated in Section~\ref{sec:corrections}. 
This also includes the estimation of the fraction of promptly produced 
$\Ds$ mesons. 
The various sources of systematic uncertainties are discussed in 
detail in Section~\ref{sec:systematics}.  
The results on the $\pt$-differential cross section compared with pQCD 
theoretical predictions, the D meson production ratios,
and the strangeness suppression factor are presented in 
Section~\ref{sec:results}. 

\section{Detector layout and data collection}
\label{sec:detector}
The ALICE detector is described in detail in~\cite{aliceJINST}.
It is composed of a central barrel, a forward muon spectrometer, 
and a set of forward detectors for triggering and event characterization.
The detectors of the central barrel are located inside a large solenoid magnet
that provides a magnetic field ${\rm B}=0.5~\mathrm{T}$, parallel to the beam 
line.

$\Ds$ mesons, and their charge conjugates, were reconstructed 
in the central rapidity region from their decays into three charged hadrons
($\KKpi$), utilizing the tracking, vertexing and particle 
identification capabilities of the central barrel detectors.

The trajectories of the decay particles were reconstructed from 
their hits in the Inner Tracking System (ITS) and in the 
Time Projection Chamber (TPC) detectors in the pseudo-rapidity range 
$|\eta|<0.8$.
The ITS~\cite{ITSalign} consists of six cylindrical layers of silicon 
detectors with radii in the range between 3.9~cm and 43.0~cm.
The two innermost layers are equipped with Silicon Pixel Detectors 
(SPD), Silicon Drift Detectors (SDD) are used in the two
intermediate layers, while the two outermost layers are composed of 
double-sided Silicon Strip Detectors (SSD).
The ITS, thanks to the high spatial resolution of the reconstructed hits,
the low material budget (on average 7.7\% of a radiation length for tracks
at $\eta=0$), 
and the small distance of the innermost layer from the beam vacuum tube,
provides the capability to detect the secondary vertices originating from 
heavy flavour decays.
For this purpose, a key role is played by the two layers of SPD detectors, 
which are located
at radial positions of 3.9 and 7.6~cm from the beam line and cover
the pseudo-rapidity ranges $|\eta|< 2.0$ and  $|\eta|< 1.4$, respectively.
The TPC~\cite{TPCpaper} provides track reconstruction
with up to 159 space points per track
in a cylindrical active volume of about 90~m$^3$.
The active volume has an inner radius of about 85 cm, an outer radius of 
about 250 cm, and an overall length along the beam direction of 500~cm.

Particle identification (PID) is provided by the measurement of the 
specific ionization energy loss, $\dEdx$, in the TPC and of the flight time
in the time-of-flight (TOF) detector.
The $\dEdx$ samples measured by the TPC are reduced, by means of a truncated
mean, to a Gaussian distribution with a resolution of 
$\sigma_{\dEdx}/(\dEdx)\approx5.5\%$~\cite{TPCpaper}.
The TOF detector is positioned at 370--399~cm from the beam 
axis and covers the full azimuth for the pseudo-rapidity range $|\eta|<0.9$.
The particle identification is based on the difference between the measured  
time-of-flight and its expected value, computed for each mass hypothesis 
from the track momentum and length. 
The overall resolution on this difference is about 160~ps and it includes
the detector intrinsic resolution, the contribution from the
electronics and the calibration, the uncertainty on the start time of the 
event (i.e. the time of the collision), and the tracking and momentum 
resolution. 
The start time of the event is defined as the weighted average between
the one estimated using the particle arrival times at the 
TOF~\cite{alicepkpi900} and the one measured by the T0 detector.
The T0 detector is composed of two arrays of Cherenkov 
counters located on either side of 
the interaction point at $+350$~cm and $-70$~cm from the nominal vertex 
position along the beam-line. 
In this analysis, the time-of-flight measurement 
provides kaon/pion separation up to a momentum of about $1.5~\gev/c$.

The data sample used for the analysis consists of
298 million minimum-bias (MB) $\pp$ collisions at $\sqrt{s}=7~\tev$, 
corresponding to an integrated luminosity $L_{\rm int} =4.8~{\rm nb}^{-1}$, 
collected during the 2010 LHC run period.
%It was collected during 5 months of running of the LHC in 2010 
%with $\pp$ collisions.
The  minimum-bias trigger was based on the information of the SPD and the 
VZERO detectors.
The VZERO detector is composed of two arrays of scintillator tiles with full 
azimuthal coverage in the pseudo-rapidity regions $2.8 < \eta < 5.1$ and 
$-3.7 < \eta < -1.7$. 
Minimum-bias collisions were triggered by requiring at least one hit
in either of the VZERO counters or in the SPD ($|\eta|<2$), in coincidence 
with the arrival of 
proton bunches from both directions. 
This trigger was estimated to be sensitive to about 87\% 
of the $\pp$ inelastic cross section~\cite{Gagliardi,MBcs1}.
It was verified by means of Monte Carlo simulations based on the PYTHIA 6.4.21 \
event generator~\cite{pythia} 
(with Perugia-0 tune~\cite{perugia0})
that the minimum-bias trigger is 100\% efficient for events containing
D mesons with $\pt>1~\gev/c$ and $|y|<0.5$~\cite{alicecharm}.
Events were further selected offline to remove the contamination from 
beam-induced background using the timing information from the VZERO and the 
correlation between the number of hits and track segments (tracklets) in the 
SPD detector.

During the $\pp$ run, the luminosity in the ALICE experiment was 
limited to 0.6--$1.2\times 10^{29}~\mathrm{cm}^{-2}\mathrm{s}^{-1}$ by 
displacing the beams in the transverse plane by 3.8 times the r.m.s.\ 
of their transverse profile, thus keeping
the probability of collision pile-up below 4\% per triggered event.
The luminous region, measured from the distribution of 
the reconstructed interaction vertices, had an r.m.s.\ width of about 4--6~cm 
along the beam direction and 35--$50~\mum$  in the transverse plane 
(the quoted ranges originate from the variations of the beam conditions during 
the data taking).
Only events with a vertex found within $\pm$10~cm from the centre
of the detector along the beam line were used for the analysis. 
This requirement selects a region  where the vertex reconstruction 
efficiency is independent of its position along the beam line and it
provides almost uniform acceptance for particles within the pseudo-rapidity 
range $|\eta|< 0.8$ for all events in the analyzed sample.
Pile-up events were identified by the presence of more than one interaction
vertex reconstructed by matching hits in the two SPD layers (tracklets).
An event was rejected from the analyzed data sample if a second interaction
vertex was found, it had at least 3 associated tracklets, and it was separated 
from the first one by more than 8 mm.
The remaining undetected pile-up is negligible for the analysis described
in this paper.

\section{$\Ds$ meson reconstruction and selection}
\label{sec:sele}

$\Ds$ mesons and their antiparticles were reconstructed in the 
decay chain $\Dstophipi$ (and its charge conjugate) followed by $\phitoKKpi$. 
The branching ratio (BR) of the chain $\DstophipitoKKpi$ is 
$2.28 \pm 0.12\%$~\cite{PDG}.
It should be noted that other $\Ds$ meson decay channels can give rise to the 
same $\KKpi$ final state.
Among them, those with larger BR are $\DstoKzerostarK$ 
and $\Dstofzeropi$, with BR into the $\KKpi$ final state of $2.63 \pm 0.13\%$ 
and $1.16 \pm 0.32\%$, respectively. 
However, as it will be discussed in the following, the selection efficiency for 
these decay modes is strongly suppressed by the cuts applied to select the 
signal candidates~\footnote{
To reduce the combinatorial background, a selection exploiting the mass
of the intermediate resonant state was applied.
Since the width of the $\phi$ peak is narrower than those of 
the $\Kzerostar$ and the $\fzero$, the decay channel through the $\phi$ 
resonance, being the one that provides the best discrimination between signal 
and background, was used in the analysis.}, and therefore the measured 
yield is dominated by the $\DstophipitoKKpi$ decays.

$\Ds$ mesons have a mean proper decay length 
$c \tau=150 \pm 2~\mum$~\cite{PDG}, which makes it possible to
resolve their decay vertex from the interaction (primary) vertex.
The analysis strategy for the extraction of the signal from the 
large combinatorial background can therefore be based on the reconstruction 
and selection of secondary vertex topologies with significant separation 
from the primary vertex.

$\Ds$ meson candidates were defined from triplets of tracks with 
proper charge sign combination. 
Tracks were selected requiring $|\eta|<0.8$, $\pt>0.4~\GeV/c$, 
a minimum of 70 associated space points in the TPC, $\chi^2/{\rm ndf}<2$ 
for the track momentum fit in the TPC, and at least 2 associated hits in the 
ITS, out of which at least one has to be in either of the two SPD layers.
For tracks that satisfy these TPC and ITS selection criteria, the 
transverse momentum resolution is better than 1\% at $\pt=1~\GeV/c$ and 
about 2\% at $\pt=10~\GeV/c$.
The resolution on the track impact parameter (i.e. the distance of 
closest approach of the track to the primary interaction vertex) in the bending 
plane ($r\phi$) is better than 75~$\mu$m for $\pt> 1$ 
GeV/$c$, well reproduced in Monte Carlo simulations~\cite{alicecharm}.
 
For each $\Ds$ candidate, in order to have an unbiased estimate of the
interaction vertex, the event primary vertex was recalculated from the 
reconstructed tracks after excluding the candidate decay tracks.
The secondary vertex was reconstructed from the decay tracks
with the same algorithm used to compute the primary vertex~\cite{alicecharm}. 
The position resolution on the $\Ds$ decay vertices was
estimated via Monte Carlo simulations to be of the order of $100~\mum$ for 
each of the three coordinates with little dependence on $\pt$.
The resolution on the position of the primary vertex depends on the event 
multiplicity: for the transverse coordinates, where the information on the 
position and spread of the luminous region is used to constrain the vertex 
fit, it ranges from $40~\mum$ in low-multiplicity events to about 10~$\mum$ 
in events with 40 charged particles per unit of rapidity. 

Candidates were then filtered by applying kinematical and topological cuts 
together with particle identification criteria. 
With the track selection described above, the acceptance in 
rapidity for D mesons drops steeply to zero for $|y|\gsim 0.5$ at low $\pt$ 
and $|y|\gsim 0.8$ at $\pt\gsim 5~\gev/c$. 
A $\pt$-dependent fiducial acceptance cut was therefore applied on the D meson
rapidity, $|y|<y_{\rm fid}(\pt)$, where $\pt$ is the $\Ds$ transverse
momentum. 
The cut value, $y_{\rm fid}(\pt)$, increases from 0.5 to 0.8 in the 
transverse momentum range $0<\pt<5~\gev/c$ according to 
a second-order polynomial function and it takes
a constant value of 0.8 for $\pt>5~\gev/c$. 

The topological selections were tuned to have a large statistical
significance of the signal, while keeping the selection efficiency as high as 
possible.
It was also checked that background fluctuations were not causing
a distortion in the signal line shape by verifying that the $\Ds$ meson mass 
and its resolution were in agreement with the Particle Data Group (PDG) value 
($1.969~\gev/c^2$~\cite{PDG}) and the simulation results, respectively.
The resulting cut values depend on the transverse momentum of the candidate.

The candidates were selected according to the decay length and 
the cosine of 
the pointing angle, $\theta_{\rm pointing}$, which is the angle between the 
reconstructed D meson 
momentum and the line connecting the primary and secondary vertex. 
The three tracks composing the candidate triplet were required to have small 
distance to the reconstructed decay vertex.
In addition, $\Ds$ candidates were selected by requiring that one
of the two pairs of opposite-charged tracks has an invariant mass compatible 
with 
the PDG world average for the $\phi$ mass (1.019~$\gev/c^2$~\cite{PDG}).
To further suppress the combinatorial background, the angles 
$\theta^*(\pi)$ and $\theta^\prime({\rm K})$ were exploited. 
$\theta^*(\pi)$ is the angle between the pion in the KK$\pi$ 
rest frame and the KK$\pi$ flight line, which is 
defined by the positions of the primary and secondary vertices 
in the laboratory frame.
$\theta^\prime({\rm K})$ is the angle between one of the kaons and the pion 
in the KK rest frame.
The cut values used for the $\Ds$ mesons with 2 $< \pt <$ 4 GeV/$c$ were: 
decay length larger than 350~$\mum$, 
$\rm cos \theta_{\rm pointing}>$ 0.94,
$|M^{\rm inv}_{\rm K^{+}K^{-}} - M^{\rm PDG}_{\phi}|<8~\MeV/c^2$,
$\cos \theta^*(\pi) < 0.95$, and 
$|\cos^3 \theta^\prime({\rm K})| > 0.1$.
A looser selection was applied at higher $\pt$ due to the lower 
combinatorial background, resulting in a selection efficiency that 
increases with increasing $\pt$.

Particle identification selections, based on the specific energy loss, 
$\dEdx$, from the TPC and the time-of-flight from the TOF detector, were used
to obtain further reduction of the background.
Compatibility cuts were applied to the difference between the measured 
signals and those expected for a pion or a kaon.
A track was considered compatible with the kaon or pion hypothesis if both
its $\dEdx$ and time-of-flight were within $3\sigma$ from the expected 
values, with at least one of them within $2\sigma$.
Tracks without a TOF signal were identified using only the TPC information 
and requiring a $2\sigma$ compatibility with the expected $\dEdx$.
Candidate triplets were required to have two tracks compatible with the kaon
hypothesis and one with the pion hypothesis.
In addition, since the decay particle with opposite charge sign has to be a 
kaon, a triplet was rejected if the opposite-sign track was not compatible 
with the kaon hypothesis.
This particle identification strategy preserves more than 90\% of the $\Ds$ 
signal and provides a reduction of the combinatorial background under the 
$\Ds$ peak by a factor of 10 in the lowest $\pt$ interval ($2<\pt<4~\gev/c$), 
a factor of 5 in $4<\pt<6~\gev/c$ and a factor of 2 at higher transverse 
momenta.

For each candidate, two values of invariant mass can be computed, corresponding
to the two possible assignments of the kaon and pion mass to the two same-sign
tracks.
Signal candidates with wrong mass assignment to the same-sign tracks
would give rise to a contribution to the invariant mass distributions that 
could potentially introduce a bias in the measured raw yield of $\Ds$ mesons.
It was verified, both in data and in simulations, that this contribution
is reduced to a negligible level by the particle 
identification selection and by the requirement that the
invariant mass of the two tracks identified as kaons is compatible with 
the $\phi$ PDG mass.

\begin{figure}[!t]
  \begin{center}
  \includegraphics[width=.99\textwidth]{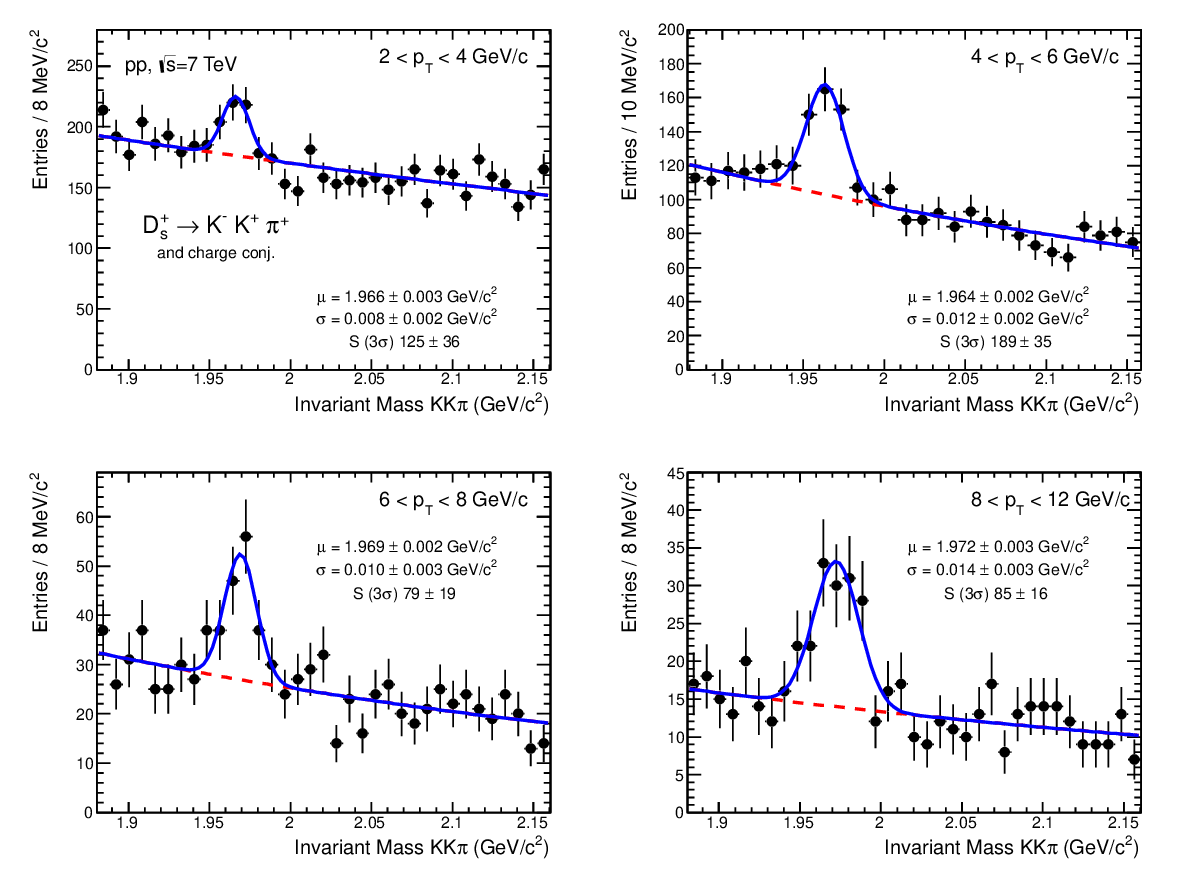}
  \caption{Invariant mass distributions for $\Ds$ candidates and charge
conjugates in the four considered $\pt$ intervals. 
The fit functions described in the text are also shown. 
The values of mean ($\mu$) and width ($\sigma$) of the signal
peak are reported together with the signal counts (S) integrated in
$\pm3\sigma$ around the centroid of the Gaussian.}
  \label{fig:invmass}
  \end{center}
\end{figure}

\begin{table}[!t]
\caption{Measured raw yields ($N^{\rm D_s^\pm~raw}$), signal (S) over background 
(B) and 
statistical significance (~S/$\sqrt{{\rm S} + {\rm B}}$~) for $\Ds$ and 
their antiparticles in the four considered $\pt$ intervals.
The estimation of the systematic uncertainty on the raw yield is described in 
Section~\ref{sec:systematics}.}
\centering
\begin{tabular}{|c|c|c|c|} 
\hline 
$\pt$ interval &$N^{\rm D_s^\pm~raw}~\pm {\rm stat.} \pm {\rm syst.}$ & S/B (3$\sigma$) & Significance (3$\sigma$) \\
 (GeV/$c$) & & &\\
\hline
\phantom{0}2--4\phantom{0} & 125$\pm$36$\pm$25 & 0.12 & 3.6 \\
\phantom{0}4--6\phantom{0} & 190$\pm$35$\pm$28 & 0.26 & 6.3 \\
\phantom{0}6--8\phantom{0} & \phantom{0}79$\pm$19$\pm$12 & 0.40 & 4.8 \\
\phantom{0}8--12 &  \phantom{0}85$\pm$16$\pm$17 & 0.58 & 5.6 \\
\hline
\end{tabular}
\label{tab:yields}
\end{table}

The raw signal yields were extracted by fitting the invariant mass distributions
in each $\pt$ interval as shown in Fig.~\ref{fig:invmass}. 
The fitting function consists of a sum of a Gaussian and an exponential 
function to describe the signal and the background, respectively. 
For all $\pt$ intervals, the invariant mass range used for the fit was 
$1.88<M^{\rm inv}_{\rm KK \pi}<2.16~\gev/c^2$, chosen in order to exclude the 
region where the background shape is affected 
by $\Dplus \to \KKpi$ decays (BR=0.265\%~\cite{PDG}) that give rise to a bump 
at the $\Dplus$ invariant mass ($1.870~\gev/c^2$~\cite{PDG}). 
The mean values of the Gaussian functions in all transverse 
momentum intervals were found to be compatible within the uncertainties 
with the PDG world average for the $\Ds$ mass.
The Gaussian widths are well reproduced in Monte Carlo simulations.
The raw yield $N^{\rm D_s^\pm~raw}$ (sum of particles and antiparticles) 
was defined as the integral of the Gaussian.
The values of $N^{\rm D_s^\pm~raw}$ are reported in Table~\ref{tab:yields} 
for the different $\pt$ intervals, together with the signal-over-background 
(S/B) ratios and the statistical significance, S/$\sqrt{{\rm S} + {\rm B}}$.
For the latter two quantities, signal (S) and background (B) were
evaluated by integrating the fit functions in $\pm3\sigma$ around the 
centroid of the Gaussian.

\section{Corrections}
\label{sec:corrections}
In order to obtain the $\pt$-differential cross section for prompt (i.e. not 
coming from weak decays of beauty mesons) ${\rm D_s^\pm}$ mesons, the raw 
yields obtained 
from the invariant mass analysis ($N^{\rm D_s^\pm~raw}$) were corrected for the 
experimental acceptance, the reconstruction and selection efficiency, 
and for the contribution to the $\Ds$ measured yield from B meson decay 
feed-down.
The production cross section of prompt $\Ds$ mesons was computed as:
\begin{equation}
  \label{eq:crosssectionD}
  \left.\frac{{\rm d}\sigma^{\rm D_s^+}}{{\rm d}\pt}\right|_{|y|<0.5}=
  \frac{1}{2}\frac{1}{\Delta y\,\Delta\pt}\frac{\left.f_{\rm prompt}\cdot N^{\rm D_s^\pm~raw}\right|_{|y|<y_{\rm fid}}}{({\rm Acc}\times\epsilon)_{\rm prompt} \cdot{\rm BR} \cdot L_{\rm int}}\,.
\end{equation}
where $\Delta\pt$ is the width of the $\pt$ interval, $\Delta y$ 
($=2\,y_{\rm fid}(\pt)$) is the width of the fiducial rapidity coverage 
(see Section~\ref{sec:sele}) and BR is the decay branching 
ratio ($2.28\%$~\cite{PDG}).
The factor $f_{\rm prompt}$ is the prompt fraction of the raw yield;
$({\rm Acc}\times\epsilon)_{\rm prompt}$ is the acceptance times efficiency of 
promptly produced $\Ds$ mesons.
The efficiency $\epsilon$ accounts for vertex reconstruction, track 
reconstruction and selection, and for $\Ds$ candidate selection
with the topological and particle identification criteria 
described in Section~\ref{sec:sele}.
The factor $1/2$ accounts for the fact that the measured raw yields 
are the sum of ${\rm D_s^+}$ and ${\rm D_s^-}$, while the cross 
section is given for particles only, neglecting the
small particle-antiparticle production asymmetry observed by 
LHCb~\cite{LHCbAsymm}.
The integrated luminosity, $L_{\rm int}=4.8~\nbinv$, was computed from 
the number of analyzed events and the cross section of 
pp collisions passing the minimum-bias trigger condition
defined in Section~\ref{sec:detector}, 
$\sigma_{\rm pp,MB}=62.2~\mb$~\cite{MBcs1,MBcs2}. 
The value of $\sigma_{\rm pp,MB}$ was derived from a van der Meer scan~\cite{vdM} 
measurement, which has an uncertainty of 3.5\%, mainly due to the 
uncertainties on the beam intensities.

The acceptance and efficiency correction factors were determined using 
$\pp$ collisions simulated with the PYTHIA 6.4.21 event 
generator~\cite{pythia} with the Perugia-0 tune~\cite{perugia0}.
Only events containing D mesons were transported through the apparatus 
(using the GEANT3 transport code~\cite{geant3}) and reconstructed.
The luminous region distribution and the conditions 
(active channels, gain, noise level, and alignment) of all the ALICE 
detectors were included in the simulations, considering also their 
evolution with time during the 2010 LHC run.

\begin{figure}[!t]
  \begin{center}
  \includegraphics[width=0.48\textwidth]{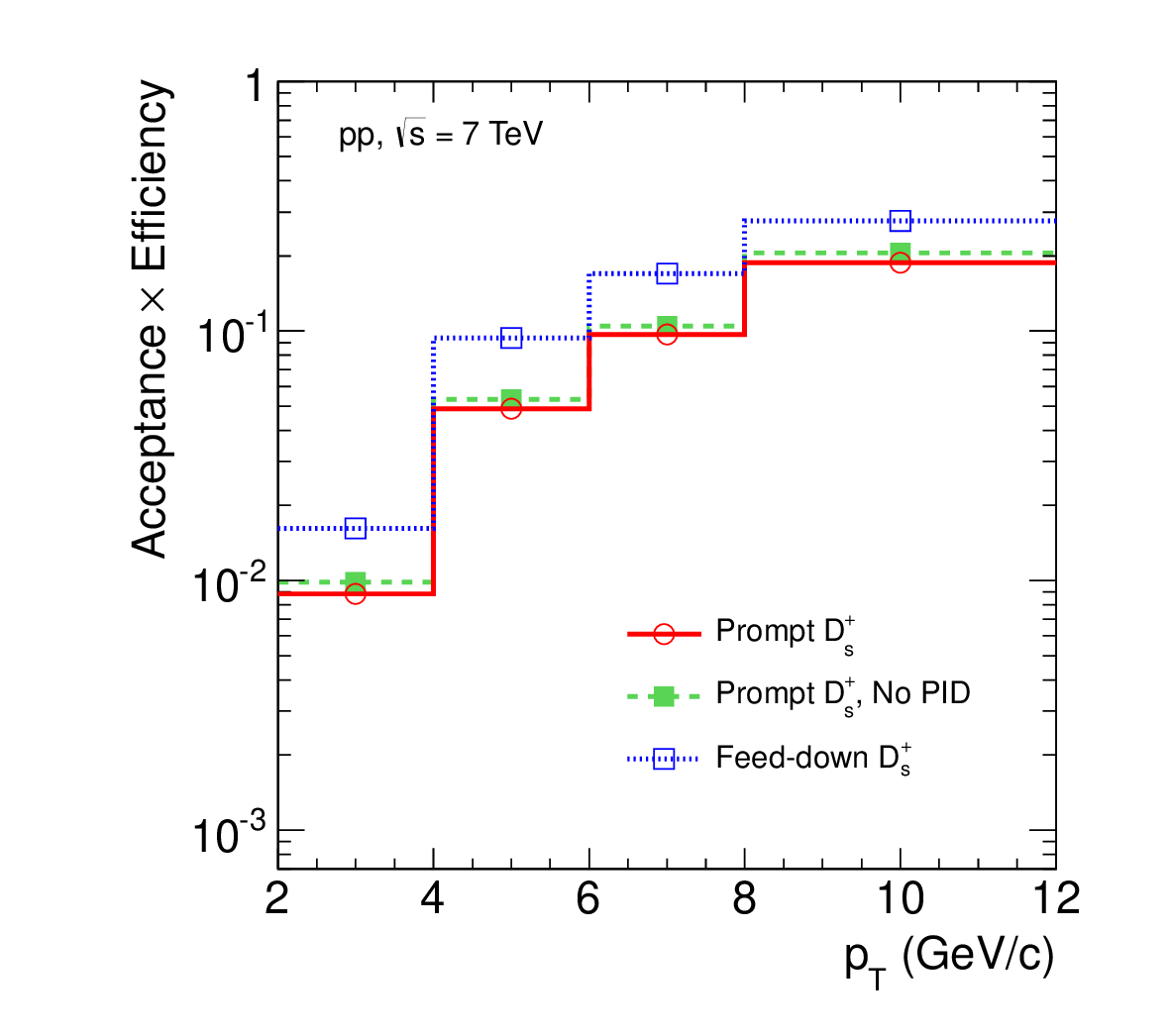}
  \includegraphics[width=0.48\textwidth]{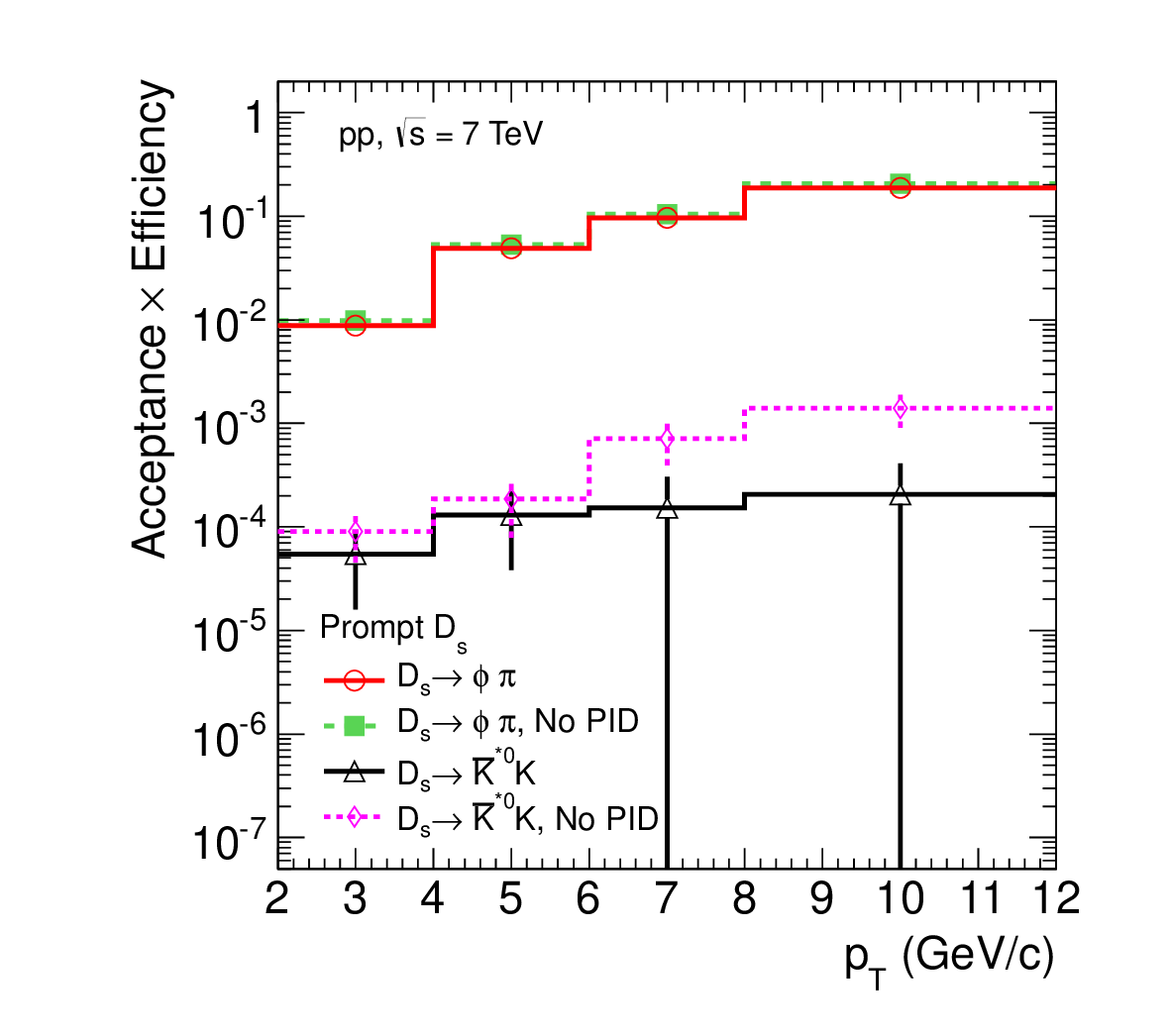}
  \caption{Acceptance $\times$ efficiency for $\Ds$ mesons as a function of 
$\pt$, for prompt and feed-down $\Ds$ mesons (left panel) and decays through 
$\phi$ and $\Kzerostar$ intermediate resonant state (right panel).}
  \label{fig:efficiencies}
  \end{center}
\end{figure}

The acceptance-times-efficiency for $\DstophipitoKKpi$ decays 
in the fiducial rapidity range described in Section~\ref{sec:sele} are shown 
in the left panel Fig.~\ref{fig:efficiencies}
for prompt and feed-down $\Ds$ mesons.
The acceptance-times-efficiency for the prompt mesons increases from about 1\%
in the lowest considered $\pt$ interval up to 10--15\% at high $\pt$.
For $\Ds$ mesons from B decays, the efficiency is larger by a factor 
1.5--2 (depending on $\pt$) because the decay vertices of the feed-down 
D mesons 
are more displaced from the primary vertex and, therefore, they are more 
efficiently selected by the topological cuts. 
The difference between the prompt and feed-down efficiencies decreases with
increasing $\pt$, because the applied selections are looser in the
higher transverse momentum intervals.
The acceptance-times-efficiency for prompt $\Ds$ mesons obtained without 
applying the particle identification selection is also shown to single out
the PID contribution to the overall efficiency.
The used particle identification strategy preserves more than 90\% of the 
signal and does not show any significant dependence on $\Ds$ meson
$\pt$ in the range considered in this analysis.

As discussed in Section~\ref{sec:sele}, the decay of the $\Ds$ meson into the 
$\KKpi$ final state occurs via different intermediate resonant states.
The selection strategy used in this analysis requires that one of the 
opposite-sign pairs of tracks composing the candidate triplet has an invariant 
mass compatible with the $\phi$ meson.
The decays $\DstophipitoKKpi$ are therefore preferentially 
selected by the applied cuts. 
Nevertheless, a fraction of the $\Ds$ decaying via another resonant state
can pass the selection cuts.
In the right panel of Fig.~\ref{fig:efficiencies}, the 
acceptance-times-efficiencies for prompt $\Ds$ decaying to $\KKpi$ final state 
via a $\phi$ and a $\Kzerostar$ in the intermediate state are 
compared. 
The acceptance-times-efficiency for the decay chain $\DstoKzerostarKtoKKpi$ is 
smaller by a factor $\approx~100$ with respect to the decay through $\phi$,
and it is further reduced when applying the PID selection.
Indeed, the PID allows the rejection of $\Ds$ decaying via a $\Kzerostar$ that
would pass the selection on the invariant mass of the $\phi$ in case of wrong
assignment of the mass (kaon/pion) to the two same-sign tracks.

The contribution to the inclusive raw yields due to $\Ds$ from B feed-down was 
subtracted using the beauty production cross section from the FONLL 
calculation~\cite{FONLL, FONLLalice},
the B$\rightarrow\Ds$ decay kinematics from the EvtGen
package~\cite{evtgen}, and the Monte Carlo efficiencies for feed-down
$\Ds$ mesons. 
Before running the EvtGen decayer, the B admixture cross section predicted by 
FONLL was split into that of $\rm B^0$, $\rm B^+$, $\rm B_s^0$ and 
$\Lambda_{\rm b}$ by assuming the same $\pt$ shape for all hadrons and the 
production fractions from~\cite{PDG}, 
namely 40.1\% of $\rm B^0$, 40.1\% of $\rm B^+$, 10.5\% of $\rm B_s^0$ and
9.3\% of beauty baryons.
The resulting fraction of prompt $\Ds$ mesons, $f_{\rm prompt}$, depends on 
the $\pt$ interval, on the applied selection cuts, and on the parameters 
used in the FONLL calculation for the B meson cross section. 
It ranges from 0.93 in the lowest transverse momentum interval 
($2<\pt<4~\gev/c$) to $\approx 0.87$ at high $\pt$ ($>6~\gev/c$).

\section{Systematic uncertainties}
\label{sec:systematics}

The systematic uncertainties on the $\Ds$ cross section 
are summarized in Table~\ref{tab:Syst} for the considered $\pt$ intervals.

The systematic uncertainty on the yield extraction was defined as the full 
spread of the $\Ds$ yield values obtained with different techniques to
analyze the invariant mass distributions in each $\pt$ interval.
The fit was repeated in different mass ranges and by varying the function 
used to describe the background.
In particular, first and second order polynomials were used instead of an 
exponential for the background. 
In case of fitting in an extended mass range, a second Gaussian signal was 
included in the fit function to account for the $\Dplus \to \KKpi$ decays. 
Furthermore, the yield extraction was repeated using a method based 
on bin counting after subtraction of the background estimated from a fit 
in the mass side bands.
The resulting uncertainty amounts to 15--20\% depending on the $\pt$ interval, 
as detailed in Table~\ref{tab:Syst}.

The systematic uncertainty on the tracking efficiency (including the effect 
of the track selection) was evaluated by comparing the probability of track 
finding in the TPC and track prolongation from the TPC to the ITS
in the data with those in the simulation, and by varying the track 
quality selections.
The estimated uncertainty is 4\% per track, which results in 12\% for the
three-body decay of $\Ds$ mesons.

Another source of systematic uncertainty originates from the residual 
discrepancies between data and simulation for the variables used to select 
the $\Ds$ candidates.
The distributions of these variables were compared for candidates passing 
loose topological cuts, i.e. essentially background candidates, and found 
to be well described in the simulation.
The effect of the imperfect implementation of the detector 
description in the Monte Carlo simulations was estimated by repeating the 
analysis with different sets of cuts.
The cut values were changed in order to vary the efficiency of signal 
selection by at least $20\%$ in all $\pt$ intervals.
A systematic uncertainty  of 15\% was estimated from the spread
of the resulting corrected yields.
Part of this uncertainty is due to residual detector misalignment effects not 
fully described in the simulation. 
To estimate this contribution, the secondary vertices in the simulation were 
reconstructed also after a track-by-track scaling of the impact parameter 
residuals with respect to their true value.
In particular, a scaling factor of 1.08, tuned to reproduce the impact 
parameter resolution observed in the data (see~\cite{alicecharm}), 
was used.
The resulting variation of the efficiency was found to be 4\% 
in the lowest $\pt$ interval used in this analysis and less than 1\% for 
$\pt>6~\gev/c$. 
This contribution was not included explicitly in the systematic uncertainty, 
because it is already accounted for in the cut variation study.

Due to the limited statistics, it was not possible to analyze separately
${\rm D_s^{+}}$ and ${\rm D_s^{-}}$ candidates to verify the absence
of biases coming from a different reconstruction efficiency for tracks with
positive and negative charge sign not properly described in the 
simulation~\footnote{
The small particle-antiparticle asymmetry reported by the LHCb 
Collaboration~\cite{LHCbAsymm} is negligible in this context.
}.
This check was carried out for other D meson species~\cite{alicecharm} 
without observing any significant difference between particle and antiparticle.

The systematic uncertainty induced by a different efficiency for particle 
identification in data and simulation was evaluated by comparing the
resulting $\pt$-differential cross section with that obtained using a
different PID approach based on 3$\sigma$ (instead of 2$\sigma$) cuts on 
TPC $\dEdx$ and time-of-flight signals, which preserves almost 100\% of the 
signal.
In addition, the PID efficiency, was estimated by comparing the reduction 
of signal yield due to the PID selection in data and in simulation, 
when the same topological cuts are applied.
Due to the limited statistical significance, this check could be 
performed in data only for $\Ds$ candidates integrated over the transverse 
momentum range $4<\pt<12~\gev/c$.
From these studies, a systematic uncertainty of 7\%, independent of 
$\pt$, was assigned to the PID selection.

The contribution to the measured yield from $\Ds$ decaying into the $\KKpi$ 
final state via other resonant channels (i.e. not via a $\phi$ meson) 
was found to be less than 1\% due to the much lower selection efficiency, as 
shown in the right panel of Fig.~\ref{fig:efficiencies} for the case of the 
decay through a $\Kzerostar$.
The contamination from other decay chains (all having smaller 
branching ratio than the two reported in Fig.~\ref{fig:efficiencies}) was 
also found to be negligible.

The effect on the selection efficiency due to the shape of the $\Ds$ $\pt$ 
spectrum used in the 
simulation was estimated from the relative difference between the Monte Carlo 
efficiencies obtained using two different $\pt$ shapes, namely those from 
PYTHIA~\cite{pythia} with Perugia-0 tune~\cite{perugia0} and from the 
FONLL pQCD calculation~\cite{FONLL,FONLLalice}.
The resulting contribution to the systematic uncertainty was found to be
3\% in the two lowest $\pt$ intervals, where the selection efficiency
is strongly $\pt$ dependent, and 2\% at higher $\pt$.

The systematic uncertainty from the subtraction of feed-down D mesons 
was estimated following the same approach as used for $\Dzero$, $\Dplus$ 
and $\Dstar$ mesons~\cite{alicecharm}.
The contribution of the FONLL perturbative uncertainties was included by 
varying the heavy-quark masses and the factorization and renormalization 
scales, $\mu_{\rm F}$ and $\mu_{\rm R}$, independently 
in the ranges $0.5<\mu_{\rm F}/\mt<2$, $0.5<\mu_{\rm R}/\mt<2$, 
with the constraint $0.5<\mu_{\rm F}/\mu_{\rm R}<2$, 
where $\mt=\sqrt{\pt^2+m_{\rm c}^2}$.
The mass of the b quark was varied within $4.5<m_{\rm b}<5~\gev/c^2$.
The uncertainty related to the B decay kinematics was estimated from
the difference between the results obtained using PYTHIA~\cite{pythia} 
instead of EvtGen~\cite{evtgen} for the particle decays and 
was found to be negligible with 
respect to the uncertainty on the B meson cross section in FONLL.
Furthermore, the prompt fraction obtained in each $\pt$ interval was compared 
with the results of a different procedure in which the FONLL
cross sections for prompt and feed-down D mesons and their respective 
Monte Carlo efficiencies are the input for evaluating the correction factor. 
Since FONLL does not have a specific prediction for $\Ds$ mesons, 
four different approaches were used to compute the $\pt$-differential
cross section of promptly produced $\Ds$.
The first two approaches used the FONLL prediction for the generic
admixture of charm hadrons and that for $\Dstar$ mesons (the $\Dstar$ 
mass being close to that of the $\Ds$) scaled with the fragmentation
fractions of charm quarks in the different hadronic species, 
$f({\rm c\rightarrow D})$, measured by ALEPH~\cite{ALEPH}.
The other two predictions for prompt $\Ds$ were computed using 
the $\pt$-differential cross section of c quarks from FONLL, the fractions
$f({\rm c\rightarrow D})$ from ALEPH~\cite{ALEPH}, and the 
fragmentation functions from~\cite{Braaten}, which have one parameter, $r$.
Two definitions were considered for the $r$ parameter:
i) $r=(m_{\rm D}-m_{\rm c})/m_{\rm D}$ ($m_{\rm D}$ and $m_{\rm c}$ being the masses 
of the considered D meson species and of the c quark, respectively) as proposed 
in~\cite{Braaten}; 
ii) $r=0.1$ for all mesons, as done in 
FONLL after fitting the analytical forms of~\cite{Braaten}
to the $\Dstar$ fragmentation function measured by ALEPH~\cite{cacciariFF}.
The ${\rm D}_{\rm s}^{*+}$ mesons produced in the c quark fragmentation were 
made to decay with PYTHIA and the resulting $\Ds$ were summed to the primary 
ones to obtain the prompt yield.
For all the four predictions used for prompt $\Ds$ cross section, the 
evaluation of $f_{\rm prompt}$ included the FONLL perturbative uncertainties 
from the variation of the factorization and renormalization scales in the 
range quoted above and of the c quark mass within $1.3<m_{\rm c}<1.7~\gev/c^2$.
The systematic uncertainty on the B feed-down was defined from the envelope 
of the resulting values of $f_{\rm prompt}$.
The resulting uncertainties in the transverse momentum intervals used in 
this analysis are about $^{+\phantom{0}5}_{-17}\%$ , as it can be seen in
Table~\ref{tab:Syst}.

Finally, the results have global systematic uncertainties due to the 
$\DstophipitoKKpi$ branching ratio (5.3\%~\cite{PDG}) and to the determination
of the cross section of pp collisions passing the minimum-bias trigger 
condition (3.5\%).

\begin{table}[tbh!]
\caption{Relative systematic uncertainties for the four considered $\pt$ 
intervals.} 
\centering
\begin{tabular}{|l|c|c|c|c|} 
\hline 
 & \multicolumn{4}{c|}{$\pt$ interval ($\gev/c$)}\\
 & 2--4 & 4--6 & 6--8 & 8--12\\
\hline
Raw yield extraction  & 20\% & 15\% & 15\% & 20\%\\
Tracking efficiency   & 12\% & 12\% & 12\% & 12\%\\
Topological selection efficiency        & 15\% & 15\% & 15\% & 15\%\\
PID efficiency        & \phantom{0}7\% & \phantom{0}7\% & \phantom{0}7\% & \phantom{0}7\%\\
MC $\pt$ shape        & \phantom{0}3\% & \phantom{0}3\% & \phantom{0}2\% & \phantom{0}2\%\\
Other resonant channels   & $<$1\% & $<$1\% & $<$1\% & $<$1\% \\[1ex] 
Feed-down from B      & $_{-18} ^{+\phantom{0}4}\%$ & $_{-17} ^{+\phantom{0}4}\%$
                      & $_{-15} ^{+\phantom{0}6}\%$ & $_{-17} ^{+\phantom{0}5}\%$\\[1ex]
\hline
Branching ratio       & \multicolumn{4}{c|}{5.3\%} \\
\hline
Normalization         & \multicolumn{4}{c|}{3.5\%}\\
\hline
\end{tabular}
\label{tab:Syst}
\end{table}

\section{Results}
\label{sec:results}
\subsection{$\pt$-differential $\Ds$ cross section and $\rm D$ meson ratios}

\begin{figure}[!t]  
\begin{center}        
\includegraphics[width=.55\textwidth]{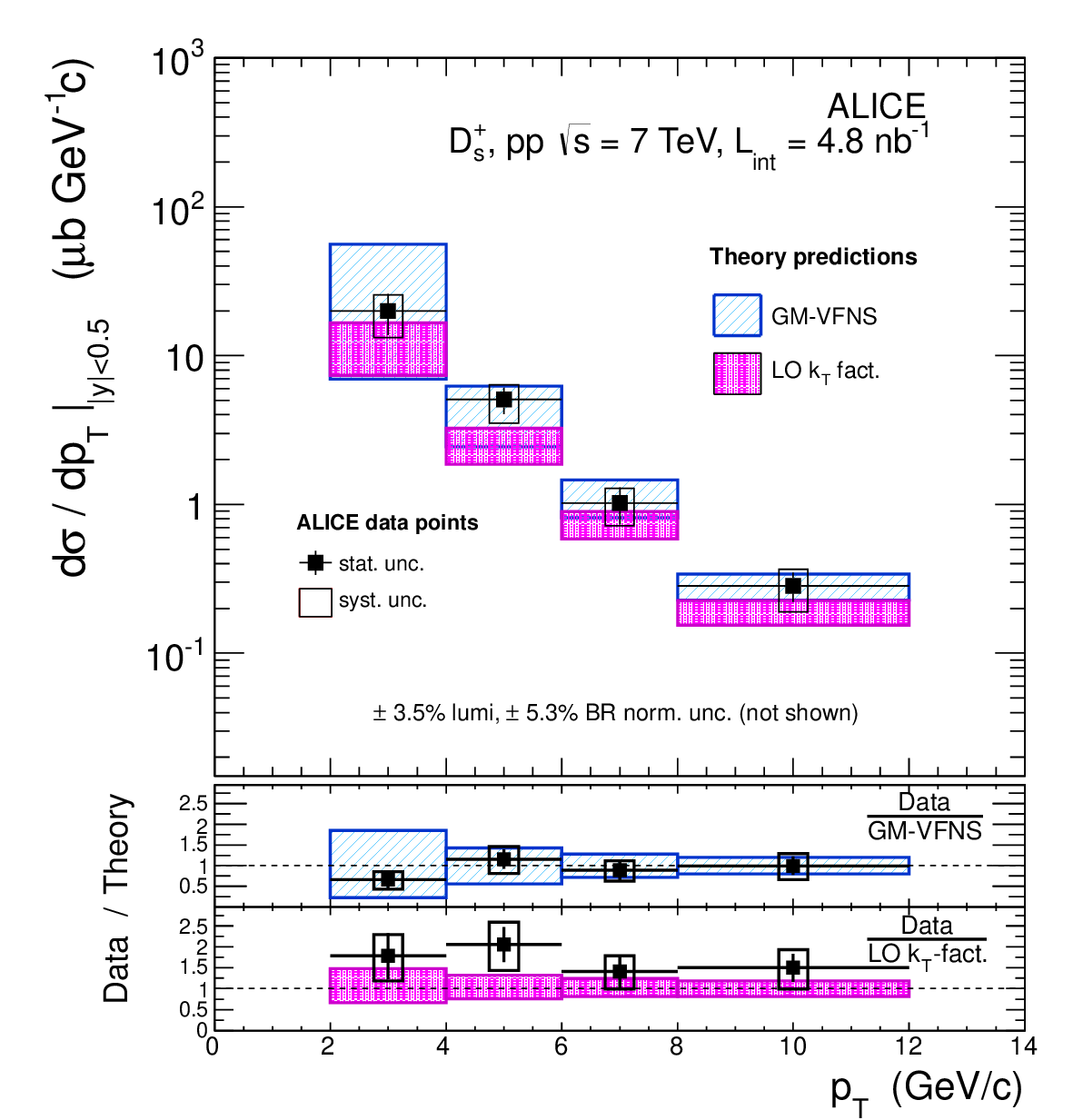}
\caption{(colour online) $\pt$-differential inclusive cross section for 
prompt $\Ds$ meson production in $\pp$ collisions at $\sqrts=7~$TeV.
The symbols are positioned horizontally at the centre of each $\pt$ interval.
The horizontal error bars represent the $\pt$ interval width. 
The normalization uncertainty (3.5\% from the minimum-bias cross 
section and 5.3\% from the branching ratio uncertainties) is not shown.
Theoretical predictions from GM-VFNS~\cite{VFNSalice} and from
$\kt$-factorization at LO~\cite{MS2,MSpriv} are also shown.}
\label{fig:ptdiffcs}
\end{center}
\end{figure}

The inclusive production cross section for prompt $\Ds$ mesons 
in four transverse momentum intervals in the range 
$2<\pt<12~\GeV/c$ is shown in Fig.~\ref{fig:ptdiffcs}. 
As discussed in section~\ref{sec:corrections}, the cross section
reported in Fig.~\ref{fig:ptdiffcs} refers to particles only, being
computed as the average of particles and antiparticles under
the assumption that the production cross section is the same for 
${\rm D_s^+}$ and ${\rm D_s^-}$.
The vertical error bars represent the statistical uncertainties, while the 
systematic uncertainties are shown as boxes around the data points.
The symbols are positioned horizontally at the centre of each $\pt$ interval,
with the horizontal bars representing the width of the $\pt$ interval. 
In Table~\ref{tab:ptdiffcs}, the numerical values of the prompt $\Ds$
production cross section are reported together 
with the average $\pt$ of $\Ds$ mesons in each transverse momentum interval.
The $\av{\pt}$ values were obtained from the $\pt$ distribution of the 
candidates in the $\Ds$ peak region, after subtracting the background 
contribution estimated from the side bands of the invariant mass distribution.
The measured differential production cross section is compared to two
theoretical predictions, namely the GM-VFNS model~\cite{gmvfnsDcdf,VFNSalice}
and the calculations from~\cite{MS2,MSpriv} based on the $\kt$-factorization approach.

The GM-VFNS prediction is found to be compatible with the measurements, 
within the uncertainties.
The central value of the GM-VFNS prediction corresponds to
the default values of the renormalization ($\mu_{\rm R}$) and factorization 
($\mu_{\rm I}$ and $\mu_{\rm F}$ for initial- and final-state singularities, 
respectively) scales, i.e. $\mu_{\rm R}=\mu_{\rm I}=\mu_{\rm F}=\mt$, 
where $\mt=\sqrt{\pt^2+m_{\rm c}^2}$, with $m_{\rm c}=1.5~\gev/c^2$.
The theoretical uncertainties are determined by varying the values of 
the renormalization and factorization scales by a factor of two up and down 
with the constraint that any ratio of the scale parameters should be
smaller than or equal to two~\cite{VFNSalice}.
The central value of the GM-VFNS prediction is higher than the
measured point by $\approx50\%$ in the first $\pt$ interval, while in the other 
intervals it agrees with the data within $\approx15\%$.
For $\Dzero$, $\Dplus$ and $\Dstar$ mesons measured by ALICE
at the same $\pp$ collision energy~\cite{alicecharm}, the central value of the 
GM-VFNS predictions was found to lie systematically above the data.
As mentioned in Section~\ref{sec:intro}, predictions for the $\Ds$ production 
cross section within the FONLL framework are not available, due to the poor 
knowledge of the fragmentation function for charm-strange mesons.

The prediction from~\cite{MS2,MSpriv} is obtained
in the framework of $\kt$-factorization at LO using
Kimber-Martin-Ryskin (KMR) unintegrated gluon distributions in the proton.
The measured $\Ds$ cross section is described by the upper limit of the 
theoretical uncertainty band.

\begin{table}[!tb]
\caption{Production cross section in $|y|<0.5$ for prompt $\Ds$ mesons in $\pp$ 
collisions at $\sqrts=7~$TeV, in $\pt$ intervals.
The normalization uncertainty (3.5\% from the minimum-bias cross 
section and 5.3\% from the branching ratio) is not included in the 
systematic uncertainties reported in the table.
The average $\pt$ of $\Ds$ mesons in each transverse momentum interval is also
reported.
}
\centering
\renewcommand{\arraystretch}{1.3}
\begin{tabular}{|c|c|c|} 
\hline 
$\pt$ interval & $\langle\pt\rangle$ & $\left.{\rm d}\sigma/{\rm d}\pt \right|_{|y|<0.5}~\pm {\rm stat.} \pm {\rm syst.}$\\
(GeV/$c$) & (GeV/$c$) & $(\mub ~ \gev^{-1} c)$\\
\hline 
\phantom{0}2--4\phantom{0} & ${ 2.7\pm0.4}$ 
       & $19.9 \pm 6.1  ^{+5.8}_{-6.7} $\\
\phantom{0}4--6\phantom{0} & ${4.7\pm0.1}$ 
       & $\phantom{0}5.06 \pm 1.03  ^{+1.3}_{-1.5} $\\
\phantom{0}6--8\phantom{0} & ${ 6.8\pm0.1}$  
       & $\phantom{0}1.02 \pm 0.28  ^{+0.27}_{-0.30} $\\
\phantom{0}8--12 & ${9.4\pm0.1}$ 
       & $0.28 \pm 0.06  ^{+0.08}_{-0.10}$ \\
\hline
\end{tabular}
\label{tab:ptdiffcs}
\end{table}

\begin{figure}[!t]  
\begin{center}        
\includegraphics[width=\textwidth]{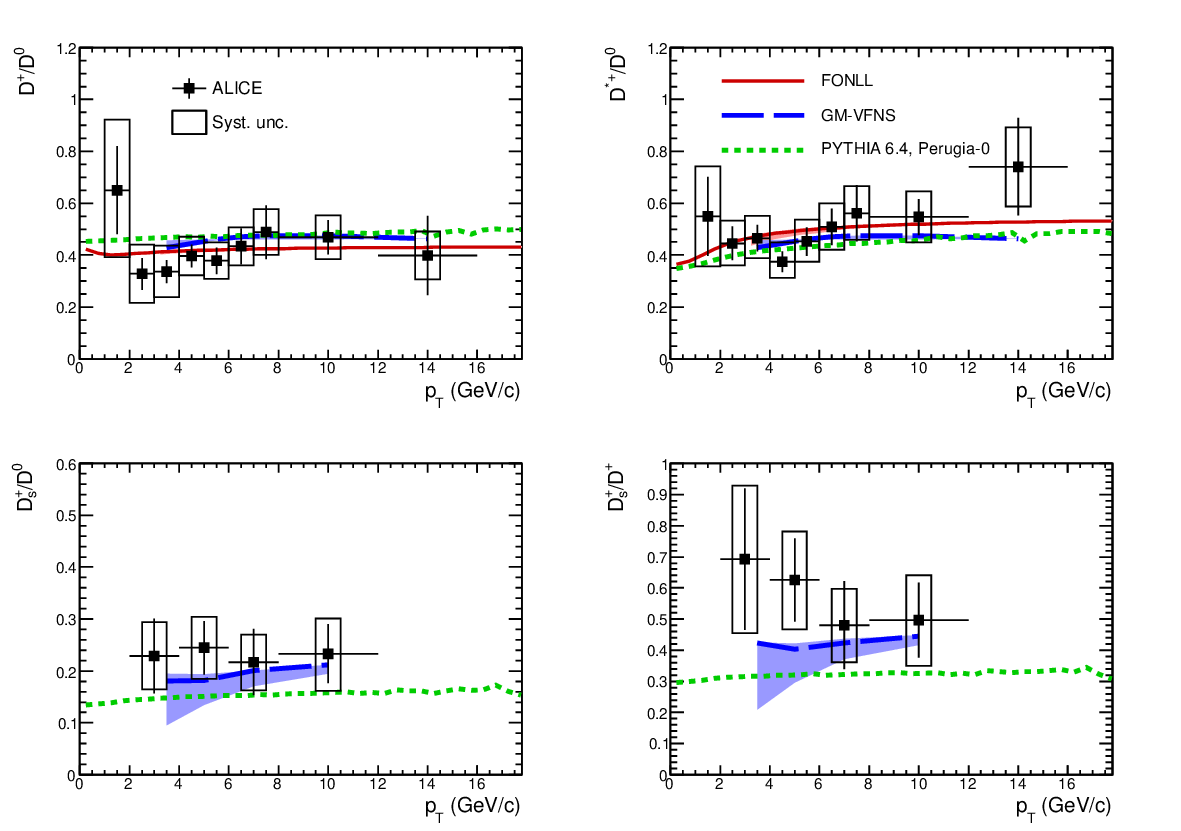}
\caption{Ratios of D meson production cross sections as a function of $\pt$.
Predictions from FONLL, GM-VFNS and PYTHIA 6.4.21 with the Perugia-0 tune 
are also shown. For FONLL and GM-VFNS the line shows the ratio of
the central values of the theoretical cross section, while the shaded area is 
defined by the ratios computed from the upper and lower limits of the 
theoretical uncertainty band.}
\label{fig:ratiosvspt}
\end{center}
\end{figure}

The ratios of the $\pt$-differential cross sections of $\Dplus$ and 
$\Dstar$ to that of $\Dzero$, taken from~\cite{alicecharm}, are shown in 
the top panels of Fig.~\ref{fig:ratiosvspt}.
In the bottom panels of the same figure, the ratios of the 
$\Ds$ cross section to the $\Dzero$ and $\Dplus$ ones are displayed.
In the evaluation of the systematic uncertainties on the D meson ratios, 
the sources of correlated and uncorrelated systematic effects were treated 
separately. 
In particular, the contributions of the yield extraction, cut efficiency and 
PID selection were considered as uncorrelated and summed in quadrature.
The systematic uncertainty on the B feed-down subtraction, being 
completely correlated, was estimated from the spread of the cross section
ratios obtained by varying the factorization and 
renormalization scales and the heavy quark mass in FONLL 
coherently for all mesons.
The uncertainty on the tracking efficiency cancels completely in the
ratios between production cross sections of mesons reconstructed from 
three-body decay channels ($\Dplus$, $\Dstar$ and $\Ds$), while a 4\% 
systematic error was considered in the ratios involving the $\Dzero$
mesons, which are reconstructed from a two-particle final state.
The $\Ds$/$\Dzero$ and $\Ds$/$\Dplus$ ratios were corrected for the different
value of pp minimum-bias cross section used in~\cite{alicecharm} and in this 
analysis~\footnote{
The preliminary pp minimum-bias cross section value of 62.5~$\mb$,
used in~\cite{alicecharm}, was updated to 62.2~$\mb$.}.

The predictions from FONLL (only for $\Dzero$, $\Dplus$ and $\Dstar$ mesons),
GM-VFNS, and the PYTHIA 6.4.21 event generator with the Perugia-0 tune
are also shown~\footnote{
The ratios from the $\kt$-factorization model of~\cite{MS2} are not shown
in Fig.~\ref{fig:ratiosvspt}. 
Indeed, in this model the fragmentation fractions $f({\rm c\rightarrow D})$
are taken from experimental measurements and the same fragmentation 
function is used for all D meson species, resulting in ratios of D meson 
cross sections that are independent of $\pt$.}.
For all these model predictions, D mesons in the rapidity range $|y|<0.5$ were 
considered.
In PYTHIA, the default configuration of the Perugia-0 tune
for charm hadronization was used.
%, except for the charm quark mass that 
%was set to 1.2~$\gev/c^2$. 
%The charm quark mass does not have an influence on the D meson production ratios
%predicted by PYTHIA.

The $\Dplus$/$\Dzero$ and $\Dstar$/$\Dzero$ ratios are determined in PYTHIA
by an input parameter, PARJ(13), that defines the probability that a charm 
or heavier meson has spin 1. 
In the Perugia-0 tune, this parameter is set to 0.54 from the measured 
fractions ${\rm P}_{\rm v}$ of heavy flavour mesons produced in vector state,
see e.g.~\cite{AndreDavid,Zeus2,alicecharmlowen}.
This setting results in an enhancement of the $\Dplus$/$\Dzero$ and a 
reduction of the $\Dstar$/$\Dzero$ ratios with respect to those obtained 
with the default value, PARJ(13)=0.75, based on spin counting.

The $\Ds/\Dzero$ and $\Ds/\Dplus$ ratios in PYTHIA are governed by
another input parameter, PARJ(2), that defines the s/u (s/d) quark suppression factor
in the fragmentation process.
In the Perugia-0 tune, PARJ(2) is set to 0.2, which gives rise to a reduced
abundance of $\Ds$ mesons with respect to the default value of 0.3.
With this parameter adjustment, PYTHIA with the Perugia-0 tune reproduces 
reasonably well the value and $\pt$ shapes of the measured ratios involving 
$\Dzero$, $\Dplus$ and $\Dstar$, while it slightly underestimates the 
abundance of $\Ds$ mesons.
The fact that PYTHIA with Perugia-0 tune underestimates the strangeness 
production was already observed at the LHC in the light flavour 
sector~\cite{ALICEhyperons,CMShyperons}.

In the Perugia 2011 tune~\cite{perugia11}, PARJ(13) is set to the same value
(0.54) as in the Perugia-0 tune, while a lower value of the 
strangeness suppression factor, PARJ(2)=0.19, is used.
This results in the same values of the Perugia-0 tune for the
$\Dplus$/$\Dzero$ and $\Dstar$/$\Dzero$ ratios, and in slightly lower values 
for the $\Ds/\Dzero$ and $\Ds/\Dplus$ ratios.

The ratios of the FONLL and GM-VFNS predictions were computed 
assuming the perturbative uncertainty to be fully correlated among the D 
meson species, i.e. using the same scales for the cross sections 
at the numerator and at the denominator.
Thus, the perturbative uncertainty cancels almost completely in the ratio,
as it can be seen in Fig.~\ref{fig:ratiosvspt} where, 
for both FONLL and GM-VFNS, the line shows the result obtained from the central 
values of the theoretical predictions, and the shaded area spans the region
between the ratios computed with the upper and lower limits of the theoretical 
uncertainty band.
The predictions from FONLL and GM-VFNS agree within uncertainties with the
measured particle ratios.
Indeed, in FONLL and GM-VFNS, the relative abundances of the various D meson 
species are not predicted by the theory: the fragmentation fractions 
$f({\rm c\rightarrow D})$ are taken from the experimental measurements.
On the other hand, in both the pQCD calculations, the $\pt$ dependence 
of the ratios of the D meson production cross sections arises from the 
different fragmentation functions used to model the transfer of energy from 
the charm quark to a specific D meson species~\cite{FONLLFF,BAKFF1,BAKFF2} 
and from the different contribution from decays of higher excited states.
The parton fragmentation models used in the calculations provide
an adequate description of the measured data.
The measured $\Ds$/$\Dzero$ and $\Ds$/$\Dplus$ ratios do not show a 
significant $\pt$ dependence within the experimental uncertainties, 
thus suggesting a small difference 
between the fragmentation functions of c quarks to strange and non-strange 
mesons.
A higher statistics data sample would be needed to conclude on a possible 
$\pt$ dependence of the ratios of strange to non-strange D meson 
cross sections.

\subsection{$\pt$-integrated $\Ds$ cross section and $\rm D$ meson ratios}

The visible cross section of prompt $\Ds$ mesons, obtained by integrating
the $\pt$-differential cross section in the measured $\pt$ range 
($2<\pt<12~\gev/c$), is
\[\sigma_{\rm vis}^{\Ds}(2<\pt<12~\gev/c,\,|y|<0.5)=\rm 53 \pm 12(stat.) _{-15}^{+13}(syst.) \pm 2(lumi.) \pm 3(BR) ~\mub.\]
The production cross section per unit of rapidity, ${\rm d}\sigma/{\rm d}y$, at 
mid-rapidity was computed by extrapolating the visible cross section
to the full $\pt$ range.
The extrapolation factor was extracted from the FONLL-based predictions for
the $\Ds$ $\pt$-differential cross section described in 
Section~\ref{sec:systematics}.
The extrapolation factor was taken as the ratio between
the total $\Ds$ production cross section in $|y|<0.5$ and the cross section 
integrated in $|y|<0.5$ and in the $\pt$ range where the experimental
measurement is performed.
In particular, the central value of the extrapolation factor was computed 
from the prediction based on the $\pt$-differential cross section of c 
quarks from FONLL, the fractions $f({\rm c\rightarrow D})$ from 
ALEPH~\cite{ALEPH}, and the fragmentation functions from~\cite{Braaten} with 
$r=0.1$.
The uncertainty on the extrapolation factor was obtained as a quadratic 
sum of the uncertainties from charm mass and perturbative scales, varied
in the ranges described above, and from the CTEQ6.6 parton distribution
functions~\cite{cteq6}.
Furthermore, to account for the uncertainty on the $\Ds$ fragmentation 
function, the extrapolation factors and their uncertainties were also computed 
using the FONLL predictions for $\Dzero$, $\Dplus$ and $\Dstar$ mesons 
and the envelope of the results was assigned as systematic uncertainty.
The resulting value for the extrapolation factor is $2.23^{+0.71}_{-0.65}$.
The prompt $\Ds$ production cross section per unit of rapidity in $|y|<0.5$ 
is then
\[{\rm d}\sigma^{\Ds}/{\rm d}y= \rm 118 \pm 28(stat.) ^{+28}_{-34}(syst.) \pm 4(lumi.) \pm 6(BR) ^{+38}_{-35}(extr.)~\mub.\]

\begin{figure}[!t]  
\begin{center}        
\includegraphics[width=.48\textwidth]{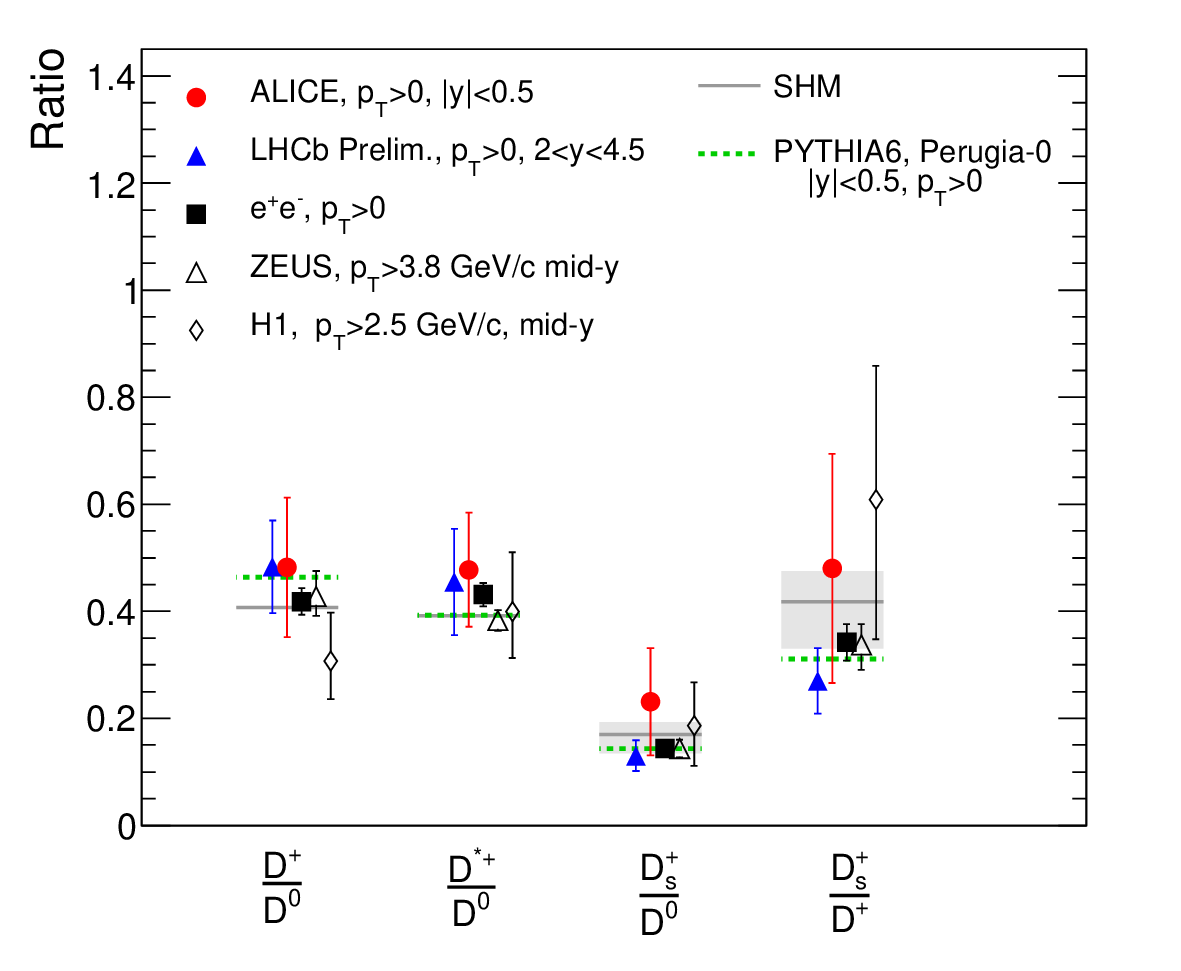}
\includegraphics[width=.48\textwidth]{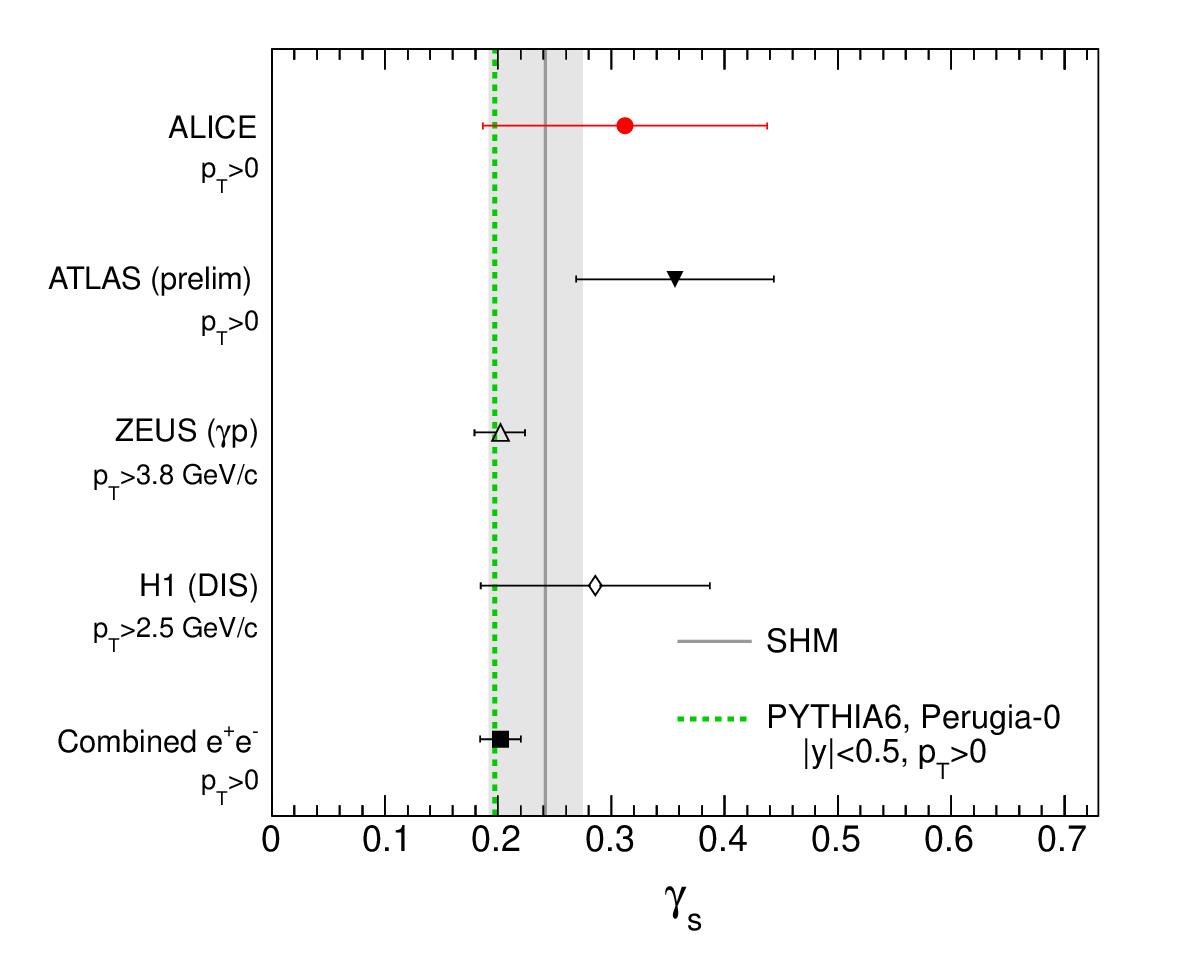}
\caption{Left: $\pt$ integrated ratios of D meson production cross sections 
compared with other experiments~\cite{lep1,Zeus2,h1,LHCbDnote}.
Error bars are the quadratic sum of statistical and systematic uncertainties,
without including the uncertainty on the BR which is common to all experiments.
Right: strangeness suppression factor $\gammas$ compared to 
measurements by other experiments~\cite{lep1,Zeus2,h1,ATLASDnote}.
Predictions from PYTHIA 6.4.21 with the Perugia-0 tune 
and from a canonical implementation of the 
statistical hadronization model (SHM)~\cite{Andronic09} are also shown.
The gray band  represents the uncertainty on the SHM predictions due to the
uncertainty on the volume and on the strangeness fugacity 
(see text for details).}
\label{fig:ratios}
\end{center}
\end{figure}

The D meson production ratios were computed from the 
cross sections per unit of rapidity, ${\rm d}\sigma/{\rm d}y$.
The corresponding values for $\Dzero$, $\Dplus$ and $\Dstar$ 
from~\cite{alicecharm} were corrected to account for the updated value of the 
pp minimum-bias cross section.
The systematic uncertainties on the ratios were computed taking into
account the correlated and uncorrelated sources as
described above.
The resulting values are reported in Table~\ref{tab:Dratios} and shown 
in the left-hand panel of Fig.~\ref{fig:ratios} together with the results
by other experiments that measured prompt charm production: 
LHCb~\cite{LHCbDnote}, 
$\epluseminus$ data (taken from the compilation in~\cite{lep1}), 
and ep data in photoproduction from ZEUS~\cite{Zeus2} and DIS from 
H1~\cite{h1}.
The error bars are the quadratic sum of statistical and systematic 
uncertainties and do not include the uncertainty on the decay branching
ratios, which are common to all experiments.
The particle ratios for ZEUS and $\epluseminus$ were computed from the 
compilation of fragmentation fractions $f({\rm c\rightarrow D})$ published 
in~\cite{Zeus2} after updating the branching ratios of the considered
decay channels to the most recent values~\cite{PDG}.
For the ZEUS data, the systematic uncertainties were propagated to the 
particle ratios by properly taking into account correlated and uncorrelated 
sources~\cite{Zeuspriv}.
For the H1 data, the D meson ratios were computed starting from the 
unconstrained values of $f({\rm c\rightarrow D})$ published in~\cite{h1}, 
taking into account the correlated part of the systematic uncertainty and 
subtracting from the quoted `theoretical' uncertainty the contribution due 
to the decay branching ratio~\cite{h1priv}. 
Also in this case, a correction was applied to account
for the updates in the branching ratios of the considered decay channels.
The ALICE results are compatible with the other measurements within 
uncertainties.

The values predicted by PYTHIA 6.4.21 with the Perugia-0 tune
are also shown in the figure, as well as those from a canonical 
implementation of the
Statistical Hadronization Model (SHM)~\cite{Andronic09}.
The values from PYTHIA were obtained by integrating the prompt D meson yields 
in the range $|y|<0.5$ and $\pt>0$.
The SHM, which computes the hadron abundances assuming that particles 
originate from a hadron gas in thermodynamical equilibrium, provides a good 
description of the measured hadron yields in heavy-ion collisions at various 
energies and centralities~\cite{Andronic08}, but it can also be applied 
to small systems like $\pp$~\cite{BecattiniHeinz,Kraus} and 
$\epluseminus$~\cite{Becattini1, Becattini2}.
The SHM results used for the present comparison were computed for
prompt D mesons, assuming a temperature $T$ of 164~MeV and a volume 
$V$ of 30$\pm$10~fm$^3$ at the moment of hadron decoupling.
The dependence on temperature of the cross section ratios considered in this
analysis is rather small within the few MeV uncertainty on the value of $T$.
To properly reproduce the yield of strange particles in small systems, 
such as $\pp$ and $\epluseminus$, an additional parameter, 
the fugacity~\cite{Kraus}, is usually introduced in the partition 
function to account for the deviation of strange particle yields from their 
chemical equilibrium values.
For the SHM predictions reported here, a value of strangeness fugacity of 
$0.60\pm0.04$, extrapolated from the results of a fit to particle yields in 
pp collisions at $\sqrts=200~\GeV$~\cite{STARthermal}, was used.
With these parameters, the SHM provides a good description of the measured 
ratios of D meson cross sections. 

\begin{table}[!t]  
\caption{Ratios of the measured production cross section for prompt D mesons 
in $\pt>0$ and $|y|<0.5$ in $\pp$ collisions at $\sqrts=7~$TeV.}
\centering
\begin{tabular}{|c|c|} 
\hline 
   & Ratio~$\pm$~(stat.)~$\pm$~(syst.)~$\pm$~(BR)\\
\hline 
$\Dplus$/$\Dzero$ & $0.48~\pm~0.07~\pm~0.11~\pm~0.01$\\
$\Dstar$/$\Dzero$ & $0.48~\pm~0.07~\pm~0.08~\pm~0.01$\\
$\Ds$/$\Dzero$ & $0.23~\pm~0.06~\pm~0.08~\pm~0.01$\\
$\Ds$/$\Dplus$ & $0.48~\pm~0.13~\pm~0.17~\pm~0.03$\\
\hline 
\end{tabular}
\label{tab:Dratios}
\end{table}

The strangeness suppression factor for charm mesons, $\gammas$, was also 
evaluated.
It is defined as 
the ratio of the production cross sections of charm-strange mesons
(c$\bar{\rm s}$) to that of non-strange charm mesons (average of 
c$\bar{\rm d}$ and c$\bar{\rm u}$)~\footnote{
The same symbol $\gammas$ is used in the statistical 
hadronization model to indicate the fugacity, which, as mentioned
above, is usually included in the partition function to account for 
strangeness suppression. 
However, the two $\gammas$ are different.
Indeed, in the statistical hadronization model, the value of the ratio between 
strange and non-strange charm mesons is proportional to the fugacity, but not
equal to it, due to the different masses of the various D meson species.}.
Since all ${\rm D^{*+}}$ and ${\rm D^{*0}}$ mesons decay into either
a $\Dzero$ or a $\Dplus$, and all ${\rm D^{*+}_{\rm s}}$ decays produce
a $\Ds$ meson~\cite{PDG}, the strangeness suppression factor
was computed as
\begin{equation}
\label{eq:strsuppeq}
\gamma_{\rm s}=\frac{2 ~{\rm d}\sigma(\Ds)/{\rm d}y}{{\rm d}\sigma(\Dzero)/{\rm d}y+{\rm d}\sigma(\Dplus)/{\rm d}y}.
\end{equation}
The contribution to $\Dzero$ and $\Dplus$ yield from decays
of excited charm-strange mesons heavier than ${\rm D^{*+}_{\rm s}}$ was 
neglected.

The resulting value of $\gammas$, computed from the $\Ds$, $\Dzero$ 
and $\Dplus$ cross sections per unit of rapidity (${\rm d}\sigma/{\rm d}y$), is
$$\gamma_{\rm s}=0.31~\pm~0.08({\rm stat.})~\pm~0.10({\rm syst.})~\pm~0.02({\rm BR}).$$
Charm-strange meson production is suppressed by a factor 
$\approx3.3$ in the fragmentation of charm quarks.
In the right-hand panel of Fig.~\ref{fig:ratios}, this result is compared with
the $\gammas$ measurements by other experiments, taken from the compilation 
in~\cite{Zeus}, after updating the branching ratios of the considered decay 
channels to the values in~\cite{PDG}.
The preliminary measurement by ATLAS~\cite{ATLASDnote} in $\pp$ 
collisions at the LHC, obtained using an equivalent 
(under the hypothesis of isospin symmetry between u and d quarks)
definition of the strangeness suppression factor based on the cross sections of 
$\Ds$, $\Dplus$ and $\Dstar$ in charm hadronization, is also shown.
The error bars are the quadratic sum of statistical and systematic 
uncertainties and do not include the uncertainty on the decay BR.
The values from PYTHIA with the Perugia-0 tune, where $\gammas$ corresponds
to PARJ(2), and the statistical hadronization model described above are also 
shown for reference.
It is also interesting to note that a similar amount of strangeness 
suppression was reported for beauty mesons by the LHCb Collaboration that
measured the ratio of strange B mesons to light neutral B mesons,
$f_{\rm s}/f_{\rm d}$, obtaining the value $0.267^{+0.021}_{-0.020}$~\cite{LHCbBfrac}.

All the $\gammas$ measurements, performed in different colliding systems and 
at different centre-of-mass energies are compatible within experimental 
uncertainties.
The current ALICE and ATLAS results at LHC energy in the central 
rapidity region do not allow one to conclude on a possible 
lifting of strangeness suppression with increasing collision energy.
Furthermore, the $\Ds$/$\Dzero$ ($\Ds$/$\Dplus$) ratios are measured at the 
LHC both at midrapidity and at forward rapidity, thus allowing to study
a possible rapidity dependence of the strangeness suppression in charm 
hadronization.
From the comparison of the ALICE and LHCb results with the current experimental
uncertainties (left-hand panel of Fig.~\ref{fig:ratios}), it is not possible 
to draw a firm conclusion on this point.

\section{Summary}
\label{sec:conclusions}

The inclusive production cross section for prompt $\Ds$ meson has been 
measured in the transverse momentum range $2<\pt<12~\gev/c$ at central 
rapidity in $\pp$ collisions at $\sqrts$ = 7~$\tev$. 
$\Ds$ mesons were reconstructed in the hadronic decay channel
$\Dstophipi$ with $\phitoKK$, and charge conjugates, using the ALICE detector.
The measured differential cross section is described within uncertainties by
the prediction from the GM-VFNS calculation, which is based on perturbative 
QCD at NLO with the collinear factorization approach, and it is compatible
with the upper side of the uncertainty band of calculations based on the
$\kt$-factorization approach at LO. 
The relative D meson production yields and the strangeness suppression factor,
$\gamma_{\rm s}=0.31~\pm~0.08({\rm stat.})~\pm~0.10({\rm syst.})~\pm~0.02({\rm BR})$, agree 
within the present experimental uncertainties with those measured by 
other experiments for different centre-of-mass energies and colliding systems. 
More precise measurements are needed to address the possible energy and 
rapidity dependence of strangeness suppression in charm hadronization.

%%%%%%%% acknowledgements
\newenvironment{acknowledgement}{\relax}{\relax}
\begin{acknowledgement}
\section*{Acknowledgements}
The ALICE collaboration would like to thank all its engineers and technicians for their invaluable contributions to the construction of the experiment and the CERN accelerator teams for the outstanding performance of the LHC complex.
The ALICE Collaboration would like to thank M.~Cacciari and H.~Spiesberger for 
providing pQCD predictions that are used for the feed-down correction and for
comparison to the measured data.
The ALICE Collaboration would like to thank R.~Maciula and A.~Szczurek for
providing the predictions based on the $\kt$-factorization approach.
The ALICE Collaboration would like to thank the Zeus and H1 collaborations, and
in particular L.~Gladilin and K.~Krueger, for providing updated values of the 
branching fractions and D meson ratios.
\\
The ALICE collaboration acknowledges the following funding agencies for their support in building and
running the ALICE detector:
 \\
Calouste Gulbenkian Foundation from Lisbon and Swiss Fonds Kidagan, Armenia;
 \\
Conselho Nacional de Desenvolvimento Cient\'{\i}fico e Tecnol\'{o}gico (CNPq), Financiadora de Estudos e Projetos (FINEP),
Funda\c{c}\~{a}o de Amparo \`{a} Pesquisa do Estado de S\~{a}o Paulo (FAPESP);
 \\
National Natural Science Foundation of China (NSFC), the Chinese Ministry of Education (CMOE)
and the Ministry of Science and Technology of China (MSTC);
 \\
Ministry of Education and Youth of the Czech Republic;
 \\
Danish Natural Science Research Council, the Carlsberg Foundation and the Danish National Research Foundation;
 \\
The European Research Council under the European Community's Seventh Framework Programme;
 \\
Helsinki Institute of Physics and the Academy of Finland;
 \\
French CNRS-IN2P3, the `Region Pays de Loire', `Region Alsace', `Region Auvergne' and CEA, France;
 \\
German BMBF and the Helmholtz Association;
\\
General Secretariat for Research and Technology, Ministry of
Development, Greece;
\\
Hungarian OTKA and National Office for Research and Technology (NKTH);
 \\
Department of Atomic Energy and Department of Science and Technology of the Government of India;
 \\
Istituto Nazionale di Fisica Nucleare (INFN) of Italy;
 \\
MEXT Grant-in-Aid for Specially Promoted Research, Ja\-pan;
 \\
Joint Institute for Nuclear Research, Dubna;
 \\
%Korea Foundation for International Cooperation of Science and Technology (KICOS);
National Research Foundation of Korea (NRF);
 \\
CONACYT, DGAPA, M\'{e}xico, ALFA-EC and the HELEN Program (High-Energy physics Latin-American--European Network);
 \\
Stichting voor Fundamenteel Onderzoek der Materie (FOM) and the Nederlandse Organisatie voor Wetenschappelijk Onderzoek (NWO), Netherlands;
 \\
Research Council of Norway (NFR);
 \\
Polish Ministry of Science and Higher Education;
 \\
National Authority for Scientific Research - NASR (Autoritatea Na\c{t}ional\u{a} pentru Cercetare \c{S}tiin\c{t}ific\u{a} - ANCS);
 \\
Federal Agency of Science of the Ministry of Education and Science of Russian Federation, International Science and
Technology Center, Russian Academy of Sciences, Russian Federal Agency of Atomic Energy, Russian Federal Agency for Science and Innovations and CERN-INTAS;
 \\
Ministry of Education of Slovakia;
 \\
Department of Science and Technology, South Africa;
 \\
CIEMAT, EELA, Ministerio de Educaci\'{o}n y Ciencia of Spain, Xunta de Galicia (Conseller\'{\i}a de Educaci\'{o}n),
CEA\-DEN, Cubaenerg\'{\i}a, Cuba, and IAEA (International Atomic Energy Agency);
 \\
Swedish Research Council (VR) and Knut $\&$ Alice Wallenberg
Foundation (KAW);
 \\
Ukraine Ministry of Education and Science;
 \\
United Kingdom Science and Technology Facilities Council (STFC);
 \\
The United States Department of Energy, the United States National
Science Foundation, the State of Texas, and the State of Ohio.
    %%%%%%% get the lates version before submitting
\end{acknowledgement}

%%
%%
%\vspace{1cm}

%%

%
\newpage
%
%\input{}               %%%%%%%%%%% put your appendices here
%
%%%%%%%%% appendix with author list
\appendix
\section{The ALICE Collaboration}
\label{app:collab}

% Collaboration: CERN-LHC-ALICE
% Generation Date is 29 Jul 2012 23:00:19 GMT
\begingroup
\small
\begin{flushleft}
B.~Abelev\Irefn{org1234}\And
J.~Adam\Irefn{org1274}\And
D.~Adamov\'{a}\Irefn{org1283}\And
A.M.~Adare\Irefn{org1260}\And
M.M.~Aggarwal\Irefn{org1157}\And
G.~Aglieri~Rinella\Irefn{org1192}\And
A.G.~Agocs\Irefn{org1143}\And
A.~Agostinelli\Irefn{org1132}\And
S.~Aguilar~Salazar\Irefn{org1247}\And
Z.~Ahammed\Irefn{org1225}
% ...
\renewcommand\labelenumi{\textsuperscript{\theenumi}~}
\section*{Affiliation notes}
\renewcommand\theenumi{\roman{enumi}}
\begin{Authlist}
\item \Adef{M.V.Lomonosov Moscow State University, D.V.Skobeltsyn Institute of Nuclear Physics, Moscow, Russia}Also at: M.V.Lomonosov Moscow State University, D.V.Skobeltsyn Institute of Nuclear Physics, Moscow, Russia
\item \Adef{University of Belgrade, Serbia}Also at: University of Belgrade, Faculty of Physics and "Vin\v{c}a" Institute of Nuclear Sciences, Belgrade, Serbia
\end{Authlist}
\section*{Collaboration Institutes}
\renewcommand\theenumi{\arabic{enumi}~}
\begin{Authlist}
\item \Idef{org1279}Benem\'{e}rita Universidad Aut\'{o}noma de Puebla, Puebla, Mexico
\item \Idef{org1220}Bogolyubov Institute for Theoretical Physics, Kiev, Ukraine
\item \Idef{org1262}Budker Institute for Nuclear Physics, Novosibirsk, Russia
\item \Idef{org1292}California Polytechnic State University, San Luis Obispo, California, United States
\item \Idef{org1329}Central China Normal University, Wuhan, China
\item \Idef{org14939}Centre de Calcul de l'IN2P3, Villeurbanne, France
\item \Idef{org1197}Centro de Aplicaciones Tecnol\'{o}gicas y Desarrollo Nuclear (CEADEN), Havana, Cuba
\item \Idef{org1242}Centro de Investigaciones Energ\'{e}ticas Medioambientales y Tecnol\'{o}gicas (CIEMAT), Madrid, Spain
\item \Idef{org1244}Centro de Investigaci\'{o}n y de Estudios Avanzados (CINVESTAV), Mexico City and M\'{e}rida, Mexico
\item \Idef{org1335}Centro Fermi -- Centro Studi e Ricerche e Museo Storico della Fisica ``Enrico Fermi'', Rome, Italy
\item \Idef{org17347}Chicago State University, Chicago, United States
\item \Idef{org1288}Commissariat \`{a} l'Energie Atomique, IRFU, Saclay, France
\item \Idef{org1294}Departamento de F\'{\i}sica de Part\'{\i}culas and IGFAE, Universidad de Santiago de Compostela, Santiago de Compostela, Spain
\item \Idef{org1106}Department of Physics Aligarh Muslim University, Aligarh, India
\item \Idef{org1121}Department of Physics and Technology, University of Bergen, Bergen, Norway
\item \Idef{org1162}Department of Physics, Ohio State University, Columbus, Ohio, United States
\item \Idef{org1300}Department of Physics, Sejong University, Seoul, South Korea
\item \Idef{org1268}Department of Physics, University of Oslo, Oslo, Norway
\item \Idef{org1132}Dipartimento di Fisica dell'Universit\`{a} and Sezione INFN, Bologna, Italy
\item \Idef{org1270}Dipartimento di Fisica dell'Universit\`{a} and Sezione INFN, Padova, Italy
\item \Idef{org1315}Dipartimento di Fisica dell'Universit\`{a} and Sezione INFN, Trieste, Italy
\item \Idef{org1145}Dipartimento di Fisica dell'Universit\`{a} and Sezione INFN, Cagliari, Italy
\item \Idef{org1312}Dipartimento di Fisica dell'Universit\`{a} and Sezione INFN, Turin, Italy
\item \Idef{org1285}Dipartimento di Fisica dell'Universit\`{a} `La Sapienza' and Sezione INFN, Rome, Italy
\item \Idef{org1154}Dipartimento di Fisica e Astronomia dell'Universit\`{a} and Sezione INFN, Catania, Italy
\item \Idef{org1290}Dipartimento di Fisica `E.R.~Caianiello' dell'Universit\`{a} and Gruppo Collegato INFN, Salerno, Italy
\item \Idef{org1103}Dipartimento di Scienze e Innovazione Tecnologica dell'Universit\`{a} del Piemonte Orientale and Gruppo Collegato INFN, Alessandria, Italy
\item \Idef{org1114}Dipartimento Interateneo di Fisica `M.~Merlin' and Sezione INFN, Bari, Italy
\item \Idef{org1237}Division of Experimental High Energy Physics, University of Lund, Lund, Sweden
\item \Idef{org1192}European Organization for Nuclear Research (CERN), Geneva, Switzerland
\item \Idef{org1227}Fachhochschule K\"{o}ln, K\"{o}ln, Germany
\item \Idef{org1122}Faculty of Engineering, Bergen University College, Bergen, Norway
\item \Idef{org1136}Faculty of Mathematics, Physics and Informatics, Comenius University, Bratislava, Slovakia
\item \Idef{org1274}Faculty of Nuclear Sciences and Physical Engineering, Czech Technical University in Prague, Prague, Czech Republic
\item \Idef{org1229}Faculty of Science, P.J.~\v{S}af\'{a}rik University, Ko\v{s}ice, Slovakia
\item \Idef{org1184}Frankfurt Institute for Advanced Studies, Johann Wolfgang Goethe-Universit\"{a}t Frankfurt, Frankfurt, Germany
\item \Idef{org1215}Gangneung-Wonju National University, Gangneung, South Korea
\item \Idef{org1212}Helsinki Institute of Physics (HIP) and University of Jyv\"{a}skyl\"{a}, Jyv\"{a}skyl\"{a}, Finland
\item \Idef{org1203}Hiroshima University, Hiroshima, Japan
\item \Idef{org1254}Indian Institute of Technology Bombay (IIT), Mumbai, India
\item \Idef{org36378}Indian Institute of Technology Indore (IIT), Indore, India
\item \Idef{org1266}Institut de Physique Nucl\'{e}aire d'Orsay (IPNO), Universit\'{e} Paris-Sud, CNRS-IN2P3, Orsay, France
\item \Idef{org1277}Institute for High Energy Physics, Protvino, Russia
\item \Idef{org1249}Institute for Nuclear Research, Academy of Sciences, Moscow, Russia
\item \Idef{org1320}Nikhef, National Institute for Subatomic Physics and Institute for Subatomic Physics of Utrecht University, Utrecht, Netherlands
\item \Idef{org1250}Institute for Theoretical and Experimental Physics, Moscow, Russia
\item \Idef{org1230}Institute of Experimental Physics, Slovak Academy of Sciences, Ko\v{s}ice, Slovakia
\item \Idef{org1127}Institute of Physics, Bhubaneswar, India
\item \Idef{org1275}Institute of Physics, Academy of Sciences of the Czech Republic, Prague, Czech Republic
\item \Idef{org1139}Institute of Space Sciences (ISS), Bucharest, Romania
\item \Idef{org27399}Institut f\"{u}r Informatik, Johann Wolfgang Goethe-Universit\"{a}t Frankfurt, Frankfurt, Germany
\item \Idef{org1185}Institut f\"{u}r Kernphysik, Johann Wolfgang Goethe-Universit\"{a}t Frankfurt, Frankfurt, Germany
\item \Idef{org1177}Institut f\"{u}r Kernphysik, Technische Universit\"{a}t Darmstadt, Darmstadt, Germany
\item \Idef{org1256}Institut f\"{u}r Kernphysik, Westf\"{a}lische Wilhelms-Universit\"{a}t M\"{u}nster, M\"{u}nster, Germany
\item \Idef{org1246}Instituto de Ciencias Nucleares, Universidad Nacional Aut\'{o}noma de M\'{e}xico, Mexico City, Mexico
\item \Idef{org1247}Instituto de F\'{\i}sica, Universidad Nacional Aut\'{o}noma de M\'{e}xico, Mexico City, Mexico
\item \Idef{org23333}Institut of Theoretical Physics, University of Wroclaw
\item \Idef{org1308}Institut Pluridisciplinaire Hubert Curien (IPHC), Universit\'{e} de Strasbourg, CNRS-IN2P3, Strasbourg, France
\item \Idef{org1182}Joint Institute for Nuclear Research (JINR), Dubna, Russia
\item \Idef{org1143}KFKI Research Institute for Particle and Nuclear Physics, Hungarian Academy of Sciences, Budapest, Hungary
\item \Idef{org1199}Kirchhoff-Institut f\"{u}r Physik, Ruprecht-Karls-Universit\"{a}t Heidelberg, Heidelberg, Germany
\item \Idef{org20954}Korea Institute of Science and Technology Information, Daejeon, South Korea
\item \Idef{org1160}Laboratoire de Physique Corpusculaire (LPC), Clermont Universit\'{e}, Universit\'{e} Blaise Pascal, CNRS--IN2P3, Clermont-Ferrand, France
\item \Idef{org1194}Laboratoire de Physique Subatomique et de Cosmologie (LPSC), Universit\'{e} Joseph Fourier, CNRS-IN2P3, Institut Polytechnique de Grenoble, Grenoble, France
\item \Idef{org1187}Laboratori Nazionali di Frascati, INFN, Frascati, Italy
\item \Idef{org1232}Laboratori Nazionali di Legnaro, INFN, Legnaro, Italy
\item \Idef{org1125}Lawrence Berkeley National Laboratory, Berkeley, California, United States
\item \Idef{org1234}Lawrence Livermore National Laboratory, Livermore, California, United States
\item \Idef{org1251}Moscow Engineering Physics Institute, Moscow, Russia
\item \Idef{org1140}National Institute for Physics and Nuclear Engineering, Bucharest, Romania
\item \Idef{org1165}Niels Bohr Institute, University of Copenhagen, Copenhagen, Denmark
\item \Idef{org1109}Nikhef, National Institute for Subatomic Physics, Amsterdam, Netherlands
\item \Idef{org1283}Nuclear Physics Institute, Academy of Sciences of the Czech Republic, \v{R}e\v{z} u Prahy, Czech Republic
\item \Idef{org1264}Oak Ridge National Laboratory, Oak Ridge, Tennessee, United States
\item \Idef{org1189}Petersburg Nuclear Physics Institute, Gatchina, Russia
\item \Idef{org1170}Physics Department, Creighton University, Omaha, Nebraska, United States
\item \Idef{org1157}Physics Department, Panjab University, Chandigarh, India
\item \Idef{org1112}Physics Department, University of Athens, Athens, Greece
\item \Idef{org1152}Physics Department, University of Cape Town, iThemba LABS, Cape Town, South Africa
\item \Idef{org1209}Physics Department, University of Jammu, Jammu, India
\item \Idef{org1207}Physics Department, University of Rajasthan, Jaipur, India
\item \Idef{org1200}Physikalisches Institut, Ruprecht-Karls-Universit\"{a}t Heidelberg, Heidelberg, Germany
\item \Idef{org1325}Purdue University, West Lafayette, Indiana, United States
\item \Idef{org1281}Pusan National University, Pusan, South Korea
\item \Idef{org1176}Research Division and ExtreMe Matter Institute EMMI, GSI Helmholtzzentrum f\"ur Schwerionenforschung, Darmstadt, Germany
\item \Idef{org1334}Rudjer Bo\v{s}kovi\'{c} Institute, Zagreb, Croatia
\item \Idef{org1298}Russian Federal Nuclear Center (VNIIEF), Sarov, Russia
\item \Idef{org1252}Russian Research Centre Kurchatov Institute, Moscow, Russia
\item \Idef{org1224}Saha Institute of Nuclear Physics, Kolkata, India
\item \Idef{org1130}School of Physics and Astronomy, University of Birmingham, Birmingham, United Kingdom
\item \Idef{org1338}Secci\'{o}n F\'{\i}sica, Departamento de Ciencias, Pontificia Universidad Cat\'{o}lica del Per\'{u}, Lima, Peru
\item \Idef{org1316}Sezione INFN, Trieste, Italy
\item \Idef{org1271}Sezione INFN, Padova, Italy
\item \Idef{org1313}Sezione INFN, Turin, Italy
\item \Idef{org1286}Sezione INFN, Rome, Italy
\item \Idef{org1146}Sezione INFN, Cagliari, Italy
\item \Idef{org1133}Sezione INFN, Bologna, Italy
\item \Idef{org1115}Sezione INFN, Bari, Italy
\item \Idef{org1155}Sezione INFN, Catania, Italy
\item \Idef{org1322}Soltan Institute for Nuclear Studies, Warsaw, Poland
\item \Idef{org36377}Nuclear Physics Group, STFC Daresbury Laboratory, Daresbury, United Kingdom
\item \Idef{org1258}SUBATECH, Ecole des Mines de Nantes, Universit\'{e} de Nantes, CNRS-IN2P3, Nantes, France
\item \Idef{org1304}Technical University of Split FESB, Split, Croatia
\item \Idef{org1168}The Henryk Niewodniczanski Institute of Nuclear Physics, Polish Academy of Sciences, Cracow, Poland
\item \Idef{org17361}The University of Texas at Austin, Physics Department, Austin, TX, United States
\item \Idef{org1173}Universidad Aut\'{o}noma de Sinaloa, Culiac\'{a}n, Mexico
\item \Idef{org1296}Universidade de S\~{a}o Paulo (USP), S\~{a}o Paulo, Brazil
\item \Idef{org1149}Universidade Estadual de Campinas (UNICAMP), Campinas, Brazil
\item \Idef{org1239}Universit\'{e} de Lyon, Universit\'{e} Lyon 1, CNRS/IN2P3, IPN-Lyon, Villeurbanne, France
\item \Idef{org1205}University of Houston, Houston, Texas, United States
\item \Idef{org20371}University of Technology and Austrian Academy of Sciences, Vienna, Austria
\item \Idef{org1222}University of Tennessee, Knoxville, Tennessee, United States
\item \Idef{org1310}University of Tokyo, Tokyo, Japan
\item \Idef{org1318}University of Tsukuba, Tsukuba, Japan
\item \Idef{org21360}Eberhard Karls Universit\"{a}t T\"{u}bingen, T\"{u}bingen, Germany
\item \Idef{org1225}Variable Energy Cyclotron Centre, Kolkata, India
\item \Idef{org1306}V.~Fock Institute for Physics, St. Petersburg State University, St. Petersburg, Russia
\item \Idef{org1323}Warsaw University of Technology, Warsaw, Poland
\item \Idef{org1179}Wayne State University, Detroit, Michigan, United States
\item \Idef{org1260}Yale University, New Haven, Connecticut, United States
\item \Idef{org1332}Yerevan Physics Institute, Yerevan, Armenia
\item \Idef{org15649}Yildiz Technical University, Istanbul, Turkey
\item \Idef{org1301}Yonsei University, Seoul, South Korea
\item \Idef{org1327}Zentrum f\"{u}r Technologietransfer und Telekommunikation (ZTT), Fachhochschule Worms, Worms, Germany
\end{Authlist}
\endgroup

  %%%%%%% get the latest version before submitting

%
%
%
\end{document}